\documentclass[sigplan,10pt,nonacm]{acmart} 

\AtBeginDocument{%
  \providecommand\BibTeX{{%
    \normalfont B\kern-0.5em{\scshape i\kern-0.25em b}\kern-0.8em\TeX}}}

\AtBeginDocument{%
    \setlength\intextsep{2pt}%
    \setlength\textfloatsep{2pt}%
    \setlength\floatsep{2pt}}

\copyrightyear{2025}
\acmYear{2025}
\setcopyright{rightsretained}
\acmConference[PPoPP '25]{The 30th ACM SIGPLAN Annual Symposium on Principles and Practice of Parallel Programming}{March 1--5, 2025}{Las Vegas, NV, USA}
\acmBooktitle{The 30th ACM SIGPLAN Annual Symposium on Principles and Practice of Parallel Programming (PPoPP '25), March 1--5, 2025, Las Vegas, NV, USA}
\acmDOI{10.1145/3710848.3710862}
\acmISBN{979-8-4007-1443-6/25/03}

\hypersetup{hidelinks=false,colorlinks=true,linkcolor=blue,urlcolor=blue,citecolor=blue,anchorcolor=blue}

\usepackage{url} 

\usepackage{tabularx} 
\usepackage{booktabs} 
\usepackage{placeins}       
\usepackage{stfloats}          

\usepackage{subfig} 

\usepackage[dvipsnames]{xcolor}

\usepackage[table]{xcolor}
\definecolor{lightgray}{gray}{0.94}
\let\oldtabular\tabular
\let\endoldtabular\endtabular
\renewenvironment{tabular}{\rowcolors{2}{white}{lightgray}\oldtabular}{\endoldtabular}

\usepackage[maxfloats=150]{morefloats}                                               
\usepackage{microtype}                                                               
\usepackage[utf8]{inputenc}   

\newcommand\Invisible[1]{                                                            
  \marginpar{\color{white}{\fontsize{.5}{.5}\selectfont #1 }}                        
}

\newcommand\Texticle{\noindent\textbf{$\blacktriangleright$}}  

\newcommand{\Exclude}[1]{}

\newcommand\Boldly[1]{\vspace{0.5 \baselineskip} \noindent \textbf{$\blacktriangleright$} \textbf{#1} \noindent}
\newcommand\BoldSection[1]{\vspace{0.20 \baselineskip} \noindent \textbf{#1} \noindent}
\definecolor{Gray95}{gray}{0.95}
\definecolor{forestgreen}{rgb}{0.13, 0.55, 0.13}

\newcommand{\remove}[1] {}

\newcommand{\AtFoot}[1]{\let\thefootnote\relax\footnotetext{{#1}}}

\usepackage{rotating} 
\newcommand{\rot}[1]{\makebox{\rotatebox{90}{#1}}}%

\newcommand{\blu}[1]{\colorbox{blue!40}{#1}}
\newcommand{\red}[1]{\colorbox{red!40}{#1}}  

\makeatletter
\newcommand{\setword}[2]{%
  \phantomsection
  #1\def\@currentlabel{\unexpanded{#1}}\label{#2}%
}
\makeatother

\makeatletter
\newcommand{\namelabel}[1]{%
  \phantomsection
  \renewcommand{\@currentlabel}{#1}
  \label{#1}
}
\makeatother

\usepackage{microtype}   
\usepackage{listings} 
\usepackage{graphicx} 
\usepackage{bold-extra} 
 
\usepackage{ifxetex}
\ifxetex
   \usepackage{fontspec}
\else
\fi

\makeatletter
\renewcommand*\verbatim@nolig@list{}
\makeatother

\usepackage[T1]{fontenc}
\usepackage[newfloat=true,frozencache=true]{minted}
\usepackage{changepage} 

\usepackage[iso,american,cleanlook]{isodate}                                         

\usepackage[export]{adjustbox} 

\usepackage{fancyhdr}                                                             
\makeatletter
\lst@Key{countblanklines}{true}[t]%
    {\lstKV@SetIf{#1}\lst@ifcountblanklines}

\lst@AddToHook{OnEmptyLine}{%
    \lst@ifnumberblanklines\else%
       \lst@ifcountblanklines\else%
         \advance\c@lstnumber-\@ne\relax%
       \fi%
    \fi}
\makeatother

\lstdefinestyle{numbers}
{numbers=left, stepnumber=1, numberstyle=\tiny, numbersep=10pt}
\lstdefinestyle{nonumbers}
{numbers=none}

\usepackage{circledtext}  
\usepackage{circlesteps} 

\usepackage{tikz} 
\usepackage[inline]{enumitem}

\newcommand*\circled[1]{\CircledTop{\footnotesize{#1}}} 

\circledtextset{boxcolor=cyan,boxtype=o,charf=\huge,boxfill = blue!60}

\pgfkeys{/csteps/inner color=yellow}
\pgfkeys{/csteps/outer color=red}
\pgfkeys{/csteps/fill color=black}

\usepackage{amsmath}
\usetikzlibrary{arrows}
\usetikzlibrary{decorations.pathreplacing}
\usetikzlibrary{calligraphy} 

\newcommand{\orcidicon}[1]{\href{https://orcid.org/#1}{\XeTeXLinkBox{\includegraphics[scale=0.06]{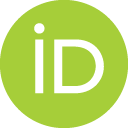}}}}

\begin{document}


\title{Reciprocating Locks} 


\author{Dave Dice \orcidicon{0000-0001-9164-7747}}                                   
\email{david.dice@gmail.com}                                                         
\orcid{0000-0001-9164-7747}                                                          
\affiliation{\institution{University of Waterloo}\country{Canada}}                                 
\author{Alex Kogan \orcidicon{0000-0002-4419-4340}}                                  
\email{alex.kogan@oracle.com}                                                        
\orcid{0000-0002-4419-4340}                                                          
\affiliation{\institution{Oracle Labs}\country{USA}}                                 
\renewcommand{\shortauthors}{Dice and Kogan}

\newcommand{\recipro}{Reciprocating Locks} 
\newcommand{\LOCKEDEMPTY}{simple locked}       
\newcommand{\Acquire}{\texttt{Acquire}}         
\newcommand{\Release}{\texttt{Release}}         

\newcommand{\nullptr}{\texttt{nullptr}}         

\newcommand\IEEEDiamondLine{\vrule depth 0pt height 0.5pt width 4cm\nobreak\hspace{7.5pt}\nobreak
  \raisebox{-3.5pt}{\fontfamily{pzd}\fontencoding{U}\fontseries{m}\fontshape{n}
  \fontsize{11}{12}\selectfont\char70}\nobreak
  \hspace{7.5pt}\nobreak\vrule depth 0pt height 0.5pt width 4cm\relax}

\begin{abstract}
We present \textbf{Reciprocating Locks}, a novel mutual exclusion locking algorithm,
targeting cache-coherent shared memory (CC), 
that enjoys a number of desirable properties.  
The \emph{doorway} arrival phase and the \Release{} operation both run in constant-time.  
Waiting threads use \emph{local spinning} and only a single \emph{waiting element} is 
required per thread, regardless of the number of locks a thread might hold at a given time.
While our lock does not provide strict FIFO admission, it bounds bypass and has 
strong anti-starvation properties.  
The lock is compact, space efficient,
and has been intentionally designed to be readily usable in real-world general purpose computing
environments such as pthreads, or C++.   
We show the lock exhibits high throughput under contention and low latency in the uncontended case. 
Under sustained contention, Reciprocating Locks generate less coherence traffic than MCS and CLH.   
The performance of \recipro{} is competitive with and often better than the best state-of-the-art scalable
queue-based spin locks.  

\end{abstract}





\renewcommand{\footruleskip}{1cm}

\maketitle

\thispagestyle{fancy} 
\fancyfoot[C]{\vspace{1cm} \textbullet \today \hspace{1mm} \textbullet \hspace{1mm}} 


\section{Introduction} 


Locks often have a crucial impact on the performance of parallel software, 
hence they remain the focus of intensive research with a 
steady stream of algorithms proposed over the last several decades. 

Reciprocating Locks was motivated by the desire for a lock algorithm that scales well 
under contention, but also avoids the various entanglements of ``queue node'' lifecycle concerns that manifest
under current state-of-the-art locks such as CLH\cite{craig-clh,ipps94-magnusson} and MCS\cite{ppopp91-Mellor-Crummey}.
Such concerns can make those locks a challenge to integrate into real-world software.  
We also wanted constant-time arrival and release paths, but were willing to forego strict FIFO admission order.
Finally, we were motivated to design an \emph{architecturally informed} lock that works well with recent developments in modern NUMA 
cache-coherent communication fabrics.


\Invisible{Oracle Patent Disclosure Accession Number IDF-139004 2024-0706 ORC25139004-US-NPR}  

Common lock algorithms include Ticket Locks, MCS and CLH.  
Ticket Locks \cite{focs79,cacm79-reed,tocs91-MellorCrummey} are simple and compact,
requiring just two words for each lock instance and no per-thread data.  
They perform well in the absence of contention, exhibiting low latency because
of short code paths.  Under contention, however, performance suffers \cite{EuroPar19-TWA} 
because all threads contending for a given lock will busy-wait on a central location, increasing
coherence costs.  For contended operation, so-called \emph{queue based} locks, such
as CLH\cite{craig-clh,ipps94-magnusson} and MCS\cite{ppopp91-Mellor-Crummey} provide relief
via \emph{local spinning} \cite{topc15-dice}.  For both CLH and MCS, arriving threads
enqueue an element (sometimes called a ``node'') onto the tail of a queue and then
busy-wait on a flag in either their own element (MCS) or the predecessor's element (CLH).  
Critically, at most one thread busy-waits on a given location at any one time, increasing
the rate at which ownership can be transferred from thread to thread relative to
techniques that use global spinning, such as Ticket Locks.

\Invisible{Alternative names : 
reciprocating; palindrome; Retrograde : apparent motion of stars ;
bifurcated; cleft or cloven; Segmented; push-swap; pop-stack; 
LIFO; FCSS = First-come soon served; boustrophedonic; push-detach; 
tmesis or tmetic lock; tomos; tomic;
} 

\Invisible{Candidate names : 
finalists :
retrograde
reciprocating

best candidates : 
Retrograde  : apparent motion of stars 
reciprocating 
bifurcated
cleft or cloven
Segmented 
push-swap 
FLIFO
FCSS = First-come soon served 
boustrophedonic
PD = Push-Detach
tmesis / tmetic lock
tomos = cut 
tomic = 
switchback

alternatives 
respiratory : inhale-exhale
two-stroke
backtracking
piston
deranged
schisma
split list 
dichotomized
cleft
bi-list
cleaved list
discursive
dual-segment or dual-list
disjointed
LIFO-FIFO
punctuated
punctual
schism 
split; fractured; partitioned
pushlock; 
alternating; toggle; up-down
yo-yo
regressive 
Violin Bow motion; sawing; sawtooth; 
ratchet
regressive
elevator 
palindrome; 
contrapuntal
ebb-flow; back-forth; to-fro; 
cancrine;
contrarian 
zig-zag
leap/jump/hop forward
walk backward then leap forward
obverse
}

\section{The Reciprocating Lock Algorithm}

Briefly, under contention, \recipro{}\footnote{PPoPP 2025 Conference Version: \url{https://doi.org/10.1145/3710848.3710862}} 
partitions the set of waiting threads into
two disjoints lists, which we call the \emph{arrival} and \emph{entry} segments.  
Threads arriving to acquire the lock will push (prepend) themselves
onto a stack, using an atomic exchange operation, forming the \emph{arrival segment}.
When the owner releases the lock, it first tries to pass ownership to
any threads found in the \emph{entry segment}.  Otherwise, if the entry segment
is found empty, the thread then uses an atomic exchange to detach the entire current arrival
segment (setting it empty), which then becomes the next entry segment, and 
then passes ownership to the first element of the entry segment 
\footnote{An implementation can be found at \url{https://github.com/pabuhr/concurrent-locking/blob/master/Reciprocating.c}}. 

In \recipro{}, a lock instance consists of an \emph{arrival} word.  We deem the lock held if  
the arrival word is non-zero.  Specifically, a value of $0$ (\texttt{nullptr}) encodes the 
\emph{unlocked} state, $1$ (\texttt{LOCKEDEMPTY}) encodes the state of \emph{simple locked} -- 
locked with an empty arrival segment -- and other 
values encode being locked, where the remainder of the arrival word points to a stack of 
threads that have recently arrived at the lock and are waiting for admission, forming the arrival segment.

Threads arriving to acquire the lock use an atomic swap (exchange) operator to install 
the address of a thread-private \emph{waiting element} into the arrival word.
If the return value from the atomic exchange was \texttt{nullptr} then the arriving
thread managed to acquire the lock without contention and can immediately enter 
the critical section.
Otherwise, our thread has encountered contention and must wait.
By virtue of the atomic exchange, the thread has managed to push its waiting element onto the arrival segment.
The non-\texttt{nullptr} value returned from the atomic exchange identifies the next thread in the stack.  
Similar to the HemLock\cite{spaa21-Dice} and CLH\cite{craig-clh,ipps94-magnusson} lock algorithms, a 
thread knows only the identity of its immediate neighbor in the arrival segment, 
and unlike MCS, no explicit linked list of waiting threads is formed or required
\footnote{Regarding terminology, say thread $A$ arrives and pushes itself onto the stack and then $B$ follows.  
In terms of arrival, $A$ is $B$’s \emph{predecessor}.  But in terms of subsequent admission order, 
in \recipro{}, $A$ is $B$’s 
\emph{successor}.  There is no such situational distinction between arrival and admission for FIFO}.  
That is, the arrival stack is implicit with no \texttt{next} pointer fields in the waiting elements. 
Our thread then proceeds to local spinning on a flag field within its waiting element.  
This flag will eventually be set during normal ownership succession by some thread running in
the \Release{} operation, passing ownership to our waiting thread.  
Our thread, still executing in \Acquire{} and now the owner, arranges to convey the address of the next thread
in the entry segment, which was obtained from the atomic exchange, to the subsequent corresponding \Release{} operation.  
The thread identified by that address will subsequently serve as the successor to our current thread.  
Our thread finally enters the critical section.

In the corresponding \Release{} operator, if a successor was passed from
the corresponding \Acquire{} operator, we simply enable that thread to enter the 
critical section by setting the flag in its waiting element.  Otherwise, we attempt to use
an atomic \texttt{compare\_and\_exchange} (CAS) operation to swing the arrival word
from \emph{simple locked} state back to \emph{unlocked}.   
If the CAS was successful then no waiting threads exist and the lock reverts to \emph{unlocked} state.
If the CAS failed, however, additional newly arrived threads
must be present on the arrival segment.  In that case we employ an atomic exchange to 
detach the entire arrival segment, leaving the arrival word in \emph{simple locked} state, encoded as $1$. 
We then pass lock ownership to the first thread in the detached segment by setting the flag in its waiting element.  


Crucially, under contention, threads arrive and join the arrival segment.  
While the entry segment remains populated, ownership is passed through the
entry segment elements in turn.
When the current entry segment becomes empty, the \Release{} operator detaches the arrival segment, 
(via an atomic exchange) which then becomes the next entry segment.
Threads migrate, in groups, from the arrival segment to the entry segment.
The arrival segment consists of those newly arrived threads currently pushed onto the 
stack anchored at the arrival word while entry segment reflects a set of threads that have 
already been detached from the arrival stack.

The \Release{} operator consults the entry segment first -- via the successor reference
passed from \Acquire{} to \Release{} --  and passes ownership to the successor 
if possible.  The sequence of successor references passed from \Acquire{} to 
\Release{} constitutes the entry segment.  
But if the entry segment is empty -- the passed successor argument is \texttt{nullptr} -- 
\Release{} then attempts to replenish the entry segment by detaching the arrival segment,
and transferring ownership to the first element.  
In the event the arrival segment is found empty, the lock reverts to \emph{unlocked} state.  


The arrival segment is implemented by means of a concurrent \emph{pop-stack}\cite{pop-stack}, where the 
key primitives are \emph{push} and \emph{detach-all}, which makes our technique immune to the \emph{A-B-A}
pathology \cite{Treiber86}.  By convention, in \recipro{}, only the current lock holder detaches the
arrival segment.  We note that Oyama et al.\cite{OTY} also use a pop-stack but they delegate execution
of the critical section.  

The waiting element is similar to the CLH or MCS ``queue node''.
In our implementation, we opt to place a thread's wait element in thread-local storage (TLS).  
As a thread can wait on at most one lock at any given time, such a singleton suffices, and
tightly bounds memory usage.  

Given that we form a stack for arriving threads, admission order is LIFO within a segment, 
but remains FIFO between segments.  As such, if thread $T1$ pushes itself onto the arrival
segment in \Acquire{}, and then waits, and $T2$ arrives and pushes itself after $T1$, 
then a given thread $T2$ can bypass or overtake $T1$ at most once before $T1$ is next granted ownership, 
providing \emph{thread-specific bounded bypass} and thus avoiding indefinite starvation.  
Alternatively, we could say \recipro{} provides classic $K$-bounded bypass (worst case) 
where $K$ reflects the cardinality of the population of threads that might compete for the lock,
yielding \emph{population bounded bypass}.  

\Invisible{
Some Confusion and variations for definition of ``bounded bypass''.
*   Wikipedia : Bounded waiting, or bounded bypass, means that the number of times a process is
    bypassed by another process after it has indicated its desire to enter the critical section
    is bounded by a function of the number of processes in the system.
*   Variations
    @   bypass by any process       -- by processes
        number of incidences of overtaking any any process
    @   bypass by any one specific  -- by a process
        number of incidences of overtaking by one process
*   terminology : bounded bypass vs bounded waiting
*   Raynal : starvation freedom
*   Keywords :
    bypass; overtake; starvation; bounded bypass; liveness; progress; overtaking
    lockout-freedom;
    finite waiting;
}

\lstloadlanguages{C++} 
\lstset{language=C++}
\lstset{frame=lines}
\lstset{basicstyle=\tiny\ttfamily} 
\lstset{morekeywords={nullptr}}   
\lstset{commentstyle=\itshape\color{gray}} 
\lstset{commentstyle=\slshape\color{gray}} 
\lstset{commentstyle=\itshape\color{gray}} 
\lstset{keywordstyle=\color{forestgreen}\bfseries} 
\lstset{backgroundcolor=\color{Gray95}}

\lstset{basicstyle=\fontsize{5.8}{6.5}\selectfont\ttfamily}
\lstset{basicstyle=\fontsize{7}{8}\selectfont\ttfamily}
\lstset{basicstyle=\fontsize{5}{6}\selectfont\ttfamily}
\lstset{basicstyle=\fontsize{6}{7}\selectfont\ttfamily}
\lstset{basicstyle=\fontsize{6.5}{7.5}\selectfont\ttfamily}
\lstset{basicstyle=\fontsize{7}{7.75}\selectfont\ttfamily}
\lstset{basicstyle=\fontsize{6}{7}\selectfont\ttfamily}
\lstset{basicstyle=\fontsize{5.4}{6}\selectfont\ttfamily}
\lstset{basicstyle=\fontsize{6}{6.5}\selectfont\ttfamily}       
\lstset{basicstyle=\fontsize{6.5}{7}\selectfont\ttfamily}       
\lstset{basicstyle=\fontsize{7}{7.5}\selectfont\ttfamily}       

\lstset{commentstyle=\color{blue}\itshape} 
\lstset{commentstyle=\itshape\rmfamily\color{blue}} 
\lstset{commentstyle=\itshape\rmfamily\color{gray}} 
\lstset{commentstyle=\itshape\sffamily\color{gray}} 
\lstset{commentstyle=\slshape\ttfamily\color{gray}} 
\lstset{commentstyle=\itshape\ttfamily\color{blue}} 
\lstset{commentstyle=\itshape\ttfamily\color{gray}}

\newcommand{\IRule}{\smash{\rule[-.2\baselineskip]{.4pt}{\baselineskip}\kern.5em}}

\section{Implementation Details}

\begin{figure}[th!]
\lstset{caption={Reciprocating Lock Algorithm}}
\lstset{label={Listing:WFvAB-lambda}}
\begin{lstlisting}[mathescape=true,escapechar=\%]
struct ReciprocatingLock { 
 %\IRule% struct WaitElement { 
 %\IRule%  %\IRule% std::atomic<WaitElement *> Gate alignas(128) {nullptr} ; 
 %\IRule% } ; 

 %\IRule% static inline const auto LOCKEDEMPTY = 
 %\IRule%  %\IRule%  (WaitElement *) uintptr_t(1) ; 

 %\IRule% // Encoding Arrivals field : 
 %\IRule% // 0:0 = Unlocked
 %\IRule% // 0:1 = Locked with empty arrival list : LOCKEDEMPTY
 %\IRule% // T:1 = Locked with populated arrival list where T is the
 %\IRule% //       most recently arrived thread on the stack
 %\IRule% std::atomic <WaitElement *> Arrivals {nullptr} ; 

 public : 
 %\IRule% inline auto %\colorbox{yellow!50}{operator+}%(std::invocable auto && csfn) %$\rightarrow$% void {
 %\IRule%  %\IRule% // %\underline{Acquire Phase}% ...
 %\IRule%  %\IRule% static thread_local WaitElement E alignas(128) {nullptr} ; 
 %\IRule%  %\IRule% E.Gate.store (nullptr, std::memory_order_release) ;
 %\IRule%  %\IRule% WaitElement * succ = nullptr ; 
 %\IRule%  %\IRule% auto eos = &E ;     // anticipate fast-path 
 %\IRule%  %\IRule% auto tail = Arrivals.exchange (&E) ; 
 %\IRule%  %\IRule% assert (tail %$\neq$% &E) ; 
 %\IRule%  %\IRule% if (tail %$\neq$% nullptr) { 
 %\IRule%  %\IRule%  %\IRule% // Coerce LOCKEDEMPTY to nullptr 
 %\IRule%  %\IRule%  %\IRule% // succ will be our successor when we subsequently release
 %\IRule%  %\IRule%  %\IRule% succ = (WaitElement *) (uintptr_t(tail) & ~1) ;
 %\IRule%  %\IRule%  %\IRule% assert (succ %$\neq$% &E) ;
 %\IRule%  %\IRule%  %\IRule%  
 %\IRule%  %\IRule%  %\IRule% // Contention -- waiting phase ...
 %\IRule%  %\IRule%  %\IRule% for (;;) {
 %\IRule%  %\IRule%  %\IRule%  %\IRule% eos = E.Gate.load(std::memory_order_acquire) ; 
 %\IRule%  %\IRule%  %\IRule%  %\IRule% if (eos %$\neq$% nullptr) break ; 
 %\IRule%  %\IRule%  %\IRule%  %\IRule% Pause() ; 
 %\IRule%  %\IRule%  %\IRule% }
 %\IRule%  %\IRule%  %\IRule% assert (eos %$\neq$% &E) ; 
 %\IRule%  %\IRule%  %\IRule%  
 %\IRule%  %\IRule%  %\IRule% // Detect logical end-of-segment sentinel marker address
 %\IRule%  %\IRule%  %\IRule% if (succ == eos) {    // terminus ? 
 %\IRule%  %\IRule%  %\IRule%  %\IRule% succ = nullptr ;    // quash !
 %\IRule%  %\IRule%  %\IRule%  %\IRule% eos  = LOCKEDEMPTY ;       
 %\IRule%  %\IRule%  %\IRule% }
 %\IRule%  %\IRule% }
 %\IRule%  %\IRule%  
 %\IRule%  %\IRule% // Critical section expressed as lambda 
 %\IRule%  %\IRule% // pass "succ" and "eos" as context 
 %\IRule%  %\IRule% assert (eos %$\neq$% nullptr) ; 
 %\IRule%  %\IRule% assert (Arrivals.load() %$\neq$% nullptr) ; 
 %\IRule%  %\IRule% csfn() ;
 %\IRule%  %\IRule% assert (Arrivals.load() %$\neq$% nullptr) ; 
 %\IRule%  %\IRule%  
 %\IRule%  %\IRule% // %\underline{Release Phase}% ...
 %\IRule%  %\IRule% // Case : entry list populated 
 %\IRule%  %\IRule% // appoint successor from entry list
 %\IRule%  %\IRule% if (succ %$\neq$% nullptr) {
 %\IRule%  %\IRule%  %\IRule% assert (eos %$\neq$% &E && succ%$\rightarrow$%Gate.load() == nullptr) ; 
 %\IRule%  %\IRule%  %\IRule% // Normal contended succession through entry segment
 %\IRule%  %\IRule%  %\IRule% // Enable successor _and propagate identity of eos
 %\IRule%  %\IRule%  %\IRule% // toward the tail of the segment 
 %\IRule%  %\IRule%  %\IRule% succ%$\rightarrow$%Gate.store (eos, std::memory_order_release) ;
 %\IRule%  %\IRule%  %\IRule% return ;
 %\IRule%  %\IRule% }
 %\IRule%  %\IRule%  
 %\IRule%  %\IRule% // Case : entry list and arrivals are both empty
 %\IRule%  %\IRule% // try fast-path uncontended unlock 
 %\IRule%  %\IRule% assert (eos == LOCKEDEMPTY || eos == &E) ; 
 %\IRule%  %\IRule% auto v = eos ; 
 %\IRule%  %\IRule% if (Arrivals.compare_exchange_strong (v, nullptr)) return ; 
 %\IRule%  %\IRule%  
 %\IRule%  %\IRule% // Case : entry list is empty and arrivals is populated 
 %\IRule%  %\IRule% // New threads have arrived and pushed themselves onto the 
 %\IRule%  %\IRule% // arrival stack. 
 %\IRule%  %\IRule% // We now detach that segment, shifting those new arrivals 
 %\IRule%  %\IRule% // to become the next entry segment 
 %\IRule%  %\IRule% auto w = Arrivals.exchange (LOCKEDEMPTY) ;
 %\IRule%  %\IRule% assert (w %$\neq$% nullptr && w %$\neq$% LOCKEDEMPTY && w %$\neq$% &E) ; 
 %\IRule%  %\IRule% assert (w%$\rightarrow$%Gate.load() == nullptr) ; 
 %\IRule%  %\IRule% w%$\rightarrow$%Gate.store (eos, std::memory_order_release) ;
 %\IRule% } 
} ; 



\end{lstlisting}  
\end{figure} 


In Listing-\ref{Listing:WFvAB-lambda} we show an implementation of \recipro{} in modern C++. 
For brevity, this version expresses the critical section as a C++ lambda expression and passes
the variables \texttt{succ} and \texttt{eos} from the acquire phase to the corresponding release phase.  
If desired, it is possible to collapse those into just one variable by a change in encoding.
Under a legacy locking interface, with distinct \texttt{lock()} and \texttt{unlock()} operators,
context may be passed via extra fields in the lock body or conveyed via thread-local storage.  
Non-escaping lambdas are efficient and add no particular runtime overhead
\footnote{Example usage : \texttt{ReciprocatingLock L \{\};\allowbreak{} int v = 5; \allowbreak{}L + [\&]\{ v += 2;\} ;}}. 
The \texttt{assert} statements in the listing express invariants and do not constitute checks against errant usage of the lock.



We also assume the existence of a ``polite'' \texttt{Pause()} operator for busy-waiting.
For encoding the arrival word, we assume that low-order bit of wait element addresses are 0.
We further assume the existence of a wait-free atomic exchange operator.  
To maintain progress properties, the 
implementation thereof should \emph{not} be via loops that employ optimistic \texttt{compare-and-swap}
or \texttt{load-locked} and \texttt{store-conditional} primitives.  In particular we 
assume that C++ \texttt{std::atomic} \texttt{exchange and compare\_and\_exchange} primitives are
implemented in a wait-free fashion, as is the case on AMD or Intel x86 processors or ARM processors
that support the LSE instruction subset.


\section{Execution Scenarios} 

\Invisible {Explicate, explain, expound, illustrate, demonstrate, show, illuminate} 
\Invisible {Narrative, annotated scenario, flow, example, depict}  
\Invisible {\recipro{} in action} 

To further explain the operation of \recipro{} we next annotate key scenarios,
showing \recipro{} in action.

\Boldly{Simple uncontended \Acquire{} and \Release{}} : 
\begin{enumerate*}[label=\protect\circled{\arabic*}]
\item Thread $T1$ arrives at Line-14 to acquire lock $L$.  $L$'s \texttt{arrival} word is currently \texttt{nullptr}, 
indicating that $L$ is in \emph{unlocked} state. 
\item At Line-17, $T1$ initializes its thread-specific waiting element, $E$, in anticipation
of potential contention.  \item $T1$ then swaps the address of $E$ 
into $L1$'s \texttt{arrival} word in Line-20.  The atomic exchange returns \texttt{nullptr}, 
indicating uncontended acquisition, so control passes into the critical section at Line-47. 
When the critical section completes execution, control enters the release phase at Line-50.  
\item As there are no successors, execution reaches the CAS operation at Line-66.  As no additional
threads have arrived, the CAS is successful and control returns from Line-66, completing the operation.  
The CAS reverts the \texttt{arrival} to \emph{unlocked} state.  
\end{enumerate*}

\Invisible{remedy, mitigate, recover, tolerate, logical, effective, tantamount}  

\Boldly{Onset of contention} In this scenario we show how to recover from a race in \Acquire{}
where a thread pushes its wait element onto the stack but, because of other arriving threads,
is not able, in the \Release{} operation, to CAS the arrival word back to $1$, 
and its element becomes ``submerged'' on the arrival stack.  We tolerate this situation by 
conveying the address of that submerged element
through the segment, during succession, allowing us to treat the buried element as the effective
end-of-segment (equivalent to \texttt{nullptr}) and otherwise ignore it.  We convey that 
address through the wait element \texttt{Gate} field, when passing ownership.
In this case we say we have a \emph{zombie} terminal element.  During succession
a thread checks the address of its successor to determine if matches the submerged terminal element. 

\begin{enumerate*}[label=\protect\circled{\arabic*}]
\item Lock $L$ is in \emph{unlocked} state and thread $T1$ arrives and acquires the lock. 
$T1$ executes the atomic exchange at Line-20 to install the address of its wait element $E$, which
we will designate $E1$, into $L$'s \texttt{arrival} word, the exchange returns $0$ (\emph{nullptr}), so $T1$ now holds the lock.
$T1$ enters the critical section.  
\item Thread $T2$ now arrives to acquire $L$ and exchanges the address of its wait element, $E2$ into 
the lock's \texttt{arrival} word.  $T2$'s successor is $T1$ and $T2$ enters the waiting loop at Line-30.
\item Thread $T3$ also arrives and pushes the address of its wait element, $E3$, onto the arrival stack.
$T3$, whose successor is $T2$, enters the waiting loop at Line-30.  
\item $T1$ finishes execution of the critical section and then starts the \Release{} operation.
At Line-66, $T1$ attempts uses a CAS to try to release $L$ replacing the address of its $E1$ 
with \emph{nullptr} encoding. 
The CAS fails, however, as other threads have arrived, and the \texttt{arrival} word now points to $E3$ instead
of $E1$.  
At this point, $T1$'s wait element $E1$ is ``buried'' or ``submerged'' in the arrival stack, 
residing at the distal end.  
\item As the CAS failed, control reaches Line-73 where $T1$ detaches the new arrivals.  
The atomic exchange replaces $E3$ with a special \texttt{LOCKEDEMPTY} distinguished value, 
which indicates the lock is held but that the arrival list is empty and has been previously detached.
\item At Line-76, $T1$ stores its \texttt{eos} (end-of-segment) address, which refers to $E1$, into $E3$'s \texttt{Gate} field.
This both enables $T3$ and conveys the address of $E1$ to $T3$.  
$T1$ has completed is locking episode.  
\item $T3$, at Line-30, observes that its \texttt{Gate} field is now $E1$. Its local \texttt{eos} variable now
refers to $E1$.  $T3$ now has ownership.  The comparison at Line-37 does not match, so $T3$ enters the critical section.
\item $T3$ executes the critcal section. 
Upon return from the critical section, $T3$ at Line-53, recognizes that it has a successor, $T2$ ($E2$),
on the detached entry segment.  At Line-58, $T3$ stores its \texttt{eos} value, $E1$, into $E2$'s \texttt{Gate} field.
The store conveys both the identity of the end-of-segment and ownership.  
$T3$ has finished its locking episode and returns at line-59.  
\item $T2$ observes at Line-30 that is has received ownership, and its \texttt{eos} value is now $E1$.  
$T2$'s indicated successor is $T1$ ($E1$) and the comparsion at Line-37 matches, so we have reached the
end of the detached entry segment. 
$T2$ then clears its local \texttt{succ} successor pointer and sets its local \texttt{eos} value to \texttt{LOCKEDEMPTY}.  
\item $T2$ enters and executes the critical section.
\item Control in $T2$ reaches line-66, where the CAS succeeds and swings the \texttt{arrival word} from \texttt{LOCKEDEMPTY}
to \emph{unlocked}.  
\end{enumerate*}

We note that a contended \Release{} operation might need to execute two atomic operations, the CAS at Line-66
and the exchange at Line-76, potentially increasing RMR complexity \cite{stoc08-attiya,podc02-anderson}. 
In practice, given the narrow window, the underlying cache line tends remain to local \emph{modified} state.
If desired, to mitigate this concern, we could condition the CAS on an immediate prior load, reducing the number of 
futile CAS operations.  We found this optimization provided no observable benefit, and did not use it. 

In the case where $E$ is submerged, we use $E$'s address for addressed-based comparisons (Line-37) 
as a distinguished marker or sentinel to indicate the logical end-of-segment.  
$E$ itself, however, will not be subsequently accessed by succession within the segment.
Elements associated with a given thread can appear on at most one segment
at any time, but, when an address is used as an end-of-segment marker, it is possible
that it appears on both the arrival segment and entry segment.

We observe that the \texttt{eos} value could also reside in a field the lock body, potentially
sequestered and isolated on a dedicated cache line, instead of 
in the waiting elements.  While viable, that approach increases the size of the lock
body, and increases induced coherence traffic.  Instead, we borrow a technique from 
Compact NUMA-Aware Locks (CNA) \cite{EuroSys19-CNA} and avoid such shared central
fields by propagating information -- in this case the address of the terminal element of the segment -- 
through the chain of waiting elements.  


\Invisible{Terminus, end-of-segment, eos, marker, sentinel, ghost, zombie, alternative end, Vesper, punctuation, virtual, phantom} 
\Invisible{end mark, tombstone, section mark; pilcrow, fleuron, hedera, dinkus, -30-} 
\Invisible{Eos; end-of-segment; terminus; Vesper; Astraeus; fini; } 
\Invisible{Boustrophedonic} 
\Invisible{UB; Nasal Demons; halt-and-catch-fire} 
\Invisible{locks diverge from academic locks}


\Boldly{Sustained contention} 
\begin{enumerate*}[label=\protect\circled{\arabic*}]
\item Lock $L$ is initially in \emph{unlocked} state.  Thread $T1$ arrives to acquire $L$.
$T1$'s exchange operation at Line-20 installs the address of $T1$'s wait element, $E1$,
into $L$'s \texttt{arrival} word.  As the exchange returned \texttt{nullptr}, $T1$ has acquired the lock.
\item $T1$ enters and executes the critical section. 
\item While $T1$ holds $L$, thread $T2$ arrives and $E2$ pushes onto the arrival stack.
The exchange operation at Line-20 returns $E1$ into $T2$'s local \texttt{tail} variable. 
As \texttt{tail} is non-$0$, $T2$ must wait at Line-29.  $T2$ local \texttt{succ} variable refers to $E1$. 
\item With $T1$ still holding $L$, $T3$ also arrives and uses the atomic exchange
to push its element $E3$ onto the arrival stack.  
The exchange returns $E2$ and $T3$'s \texttt{succ} variable points to $E2$.
The arrival stack consists of $E3$ followed by $E2$ followed by $E1$, although $E1$ is ``buried''
and will be excised during subsequent succession.  
The entry segment is empty.  
$T3$ waits on $E3$ at Line-30
\item Similarly, $T4$ arrives and pushes $E4$ onto the arrival stack and then waits.  
$T4$'s successor variable points to $E3$.  
\item $T1$ eventually completes the critical section and then executes the \Release{}
phase at Line-50.  As $T1$'s \texttt{succ} variable is \texttt{nullptr}, indicating an
empty entry segment, $T1$ then attempts the CAS, at Line-66 which fails.  
$T1$ then executes exchange(\texttt{LOCKEDEMPTY}) to detach the arrival segment at Line-73.  
The exchange operator returns $E4$.  The arrival segment is now empty and the entry segment
consists of $E4$ then $E3$ then $E2$ then $E1$.  $T1$ passes ownership to $T4$ at Line-76,
passing the address of $E1$ as the end-of-segment marker.  
\item $T4$ is now the owner and departs its waiting phase at Line-31, 
having received $E1$ as the end-of-segment address.  
The end-of-segment address check at Line-37 does not match, so $T4$ enters the critical section. 
\item while $T4$ holds $L$, $T5$ and then $T6$ arrive to acquire $L$, pushing $E5$ and then $E6$, respectively,
onto the arrival stack.  The arrival segment consists of $E6$ then $E5$ and the detached entry segment is 
just $E3$ and $E2$ and $E1$.  When $T5$ arrived, its exchange replaced \texttt{LOCKEDEMPTY} with $E5$.
Crucially, Line-25 converted \texttt{LOCKEDEMPTY}, which is encoded as $1$, to a effective
successor value of \emph{nullptr}.  
\item $T4$ releases the lock.  As $E3$ is $T4$'s the successor, we grant ownership to $T3$ at Line-58,
again passing $E1$ as the end-of-segment address.
\item $T3$ departs its wait phase at Line-31 and enters and executes the critical section.  
\item $T3$ releases the lock, observing $E2$ as its successor and grants ownership to $T2$ at Line-58,
again, passing $E1$ as the end-of-segment,  
\item $T2$ departs its wait phase at line-31.  
In this particular step, the address check at Line-37 matches, and $T2$ annuls its \texttt{succ} variable
and sets its \texttt{eos} to \texttt{LOCKEMPTY}.  
We have reached the end of the entry segment, which is now exhausted. 
$T2$ enters and executes the critical section. 
\item $T2$ executes the release phase, and since \texttt{succ} is \emph{nullptr} at Line-53, the
current entry segment is empty.  
$T2$ then attempts the CAS at line 66, which tries to replace \texttt{LOCKEDEMPTY} with \emph{unlocked}. 
The CAS fails as the arrival segment is populated with $T6$ and $T5$ and the \texttt{arrival} word
points to $E6$.  
$T2$ then detaches the new arrivals at Line-73 and passes ownership to $T6$ (via $E6$) at Line-76, and also
conveying the end-of-segment address, which is now \texttt{LOCKEDEMPTY}, to $T6$.  
\item $T6$ exits the waiting loop at Line-30, noting that \texttt{eos} is now \texttt{LOCKEDEMPTY}.
$T6$'s successor is $T5$.  $T6$ enters the critical section.  
\item $T6$ completes the critical section and executes the release phase.  As $T6$'s \texttt{succ}
variable refers to $T5$, we pass ownership and the end-of-segment address (\texttt{LOCKEDEMPTY} at
this juncture) to $T5$.  
\item $T5$ recognizes ownership at Line-30 and then enters the critical section at Line-47.
\item $T5$ releases the lock.  It's \texttt{succ} value is \emph{nullptr} to it the
attempts the CAS at Line-66.  In this case the CAS succeeds, replacing \texttt{LOCKEDEMPTY} with 
the \emph{unlocked} encoding of \emph{nullptr}.   
$L$ is restored to \emph{unlocked} state.  
\end{enumerate*} 


\section{Additional Requirements}

\Invisible{practical; pragmatic; de facto; QoI = quality of implementation, desiderata} 

We target lock algorithms that are suitable for environments such as the Linux kernel,
as a replacement for the user-level \texttt{pthreads\_mutex} primitive, or for use
in runtime environments such as the HotSpot Java Virtual Machine (JVM), which is written in C++. 
As such, we identify additional de-facto requirements for general purpose lock algorithms.

\BoldSection{Safe Against Prompt Lock Destruction} Various lock algorithms perform stores in
the \Release{} operation that potentially release ownership, but then perform additional
accesses to fields in the lock body to ensure succession and progress.  
This can result in a class of unsafe use-after-free
memory reference errors \cite{LifecycleSurprise1,LifecycleSurprise2}
if a lock is used to protect its own existence.  
A detailed description of such an error can be found in \cite{spaa21-Dice}.  
All algorithms for use in the linux kernel or in the pthreads environment are
expected to be prompt lock destruction safe, constituting a de facto requirement.

\BoldSection{Support for Large Numbers of Extant Threads} The algorithms must support the
simultaneous participation of an arbitrary number of threads, where threads
are created and destroyed dynamically, and the peak number of extant threads
is not known in advance.

\BoldSection{Support for Large Numbers of Extant Locks} Similarly, the algorithm 
needs to support large numbers of lock instances, which can be created and
destroyed dynamically.  Recent work \cite{eurosys19-lochmann} shows the 
Linux kernel has more than 6000 statically initialized lock instances.  Again, the
number of locks is not know in advance, as drivers load, and unload, for instance,
or as data structures containing locks dynamically resize. 

\Invisible{multitudes; plural locking} 

\BoldSection{Plural Locking} A given thread is expected to be able to lock
and hold a large number of locks simultaneously.
In the Linux kernel, for instance, situations arise when 40 or more distinct
locks are held by a single thread at a given time, as evidenced by the
\texttt{MAX\_LOCK\_DEPTH} tunable, which is used by the kernel's ``lockdep''
facility to track the set of locks held in an explicit per-thread list for
the purposes of detecting potential deadlock.
Furthermore, locks must be able to be released in non-LIFO imbalanced order as it
is fairly common to acquire a lock in one routine, return, and then
release the lock in the caller.  


\BoldSection{Space Efficient} We expect the lock algorithm to be frugal
and parsimonious with regard to space usage.  Critically, any algorithms
that require per-lock and per-thread storage -- with $Threads * Locks$ space consumption --
such as Anderson's Array-Based Queue Lock \cite{tpds90-Anderson}, 
are not suitable.  

\BoldSection{FIFO as a non-goal} 
We also note that many real-world lock implementations, such as the current implementation
of Java's ``synchronized'' in the HotSpot Java Virtual Machine, the \texttt{java\-.util\-.Concurrent\-.ReentrantLock},
and the default \texttt{pthread\_mutex} on Linux,
are non-FIFO, and in fact permit unbounded bypass and indefinite starvation.  
Relatedly, Compact NUMA-Aware Locks (CNA) \cite{EuroSys19-CNA}  intentionally imposes non-FIFO admission order
to improve throughput on NUMA platforms, trading aggregate throughput against strict FIFO ordering.
\footnote{Even if a lock has strict FIFO admission order, otherwise
identical threads can still suffer from persistent long-term unfairness related to
shared cache residency imbalance, reflecting the ``Matthew Effect''\cite{spaa14-dice}.}.  
In fact, strict FIFO locking is an anti-pattern for common practical 
lock designs, as it suffers from reduced throughput compared to more relaxed admission schedules,
and has shortcomings when used with waiting techniques that deschedule threads, or when
involuntary preemption is in play \cite{arxiv-Malthusian, EuroPar19-GCR, dice2020fissile, netys20-dice}.
Forgoing strict FIFO admission schedules allows more opportunism that in some circumstances may translate 
into improved throughput. 

\BoldSection{Work Conserving} We desire that the lock is work conserving and avoids the use of 
backoff delays, which can otherwise constitute ``dead time''.  



Given these particular constraints, we exclude a number of algorithms from consideration.
Dvir's algorithms \cite{opodis17-dvir}, Rhee's algorithm \cite{rhee96}, 
Lee's HL1 and HL2 \cite{Lee-HL1-TwoFace,icdcs05-lee}, 
and Jayanati's ``Ideal Queued Spinlock'' \cite{icdcn20-jayanti} \cite{Jayanti-Dissertation} algorithms,
when simplified for use in cache-coherent (CC) environments, all have extremely simple and elegant paths and
low RMR complexity, suggesting they would be competitive and good candidates, but they 
do not readily tolerate multiple locks being held simultaneously.  
The above tend to use so-called ``node-toggling'' and ``node-switching'' techniques -- also called ``two-face'' by Lee --
that, while they work well when a thread holds at most one lock, are awkward for general purpose use.
Our specific concerns are space blow-up due to the need to maintain toggle element pairs and a ``face'' index for
pairs of locks and threads, in addition to the requirement that we re-associate that metadata 
with the lock instance at release-time. 
Some of Dvir and Lee's algorithms are also not safe against prompt lock destruction.


\section{Related Work} 

While mutual exclusion remains an active research topic
\cite{ppopp17-Ramalhete,craig-clh,ppopp91-Mellor-Crummey,
EuroPar19-TWA,icdcn20-jayanti,Jayanti-Dissertation,
EuroPar19-GCR,opodis17-dvir,
EuroSys19-CNA,
eurosys17-dice,oopsla99-agesen,topc15-dice,ScottB24,MCSH,isca10-eyerman,aksenov-CLH,aksenov-MCS,tocs19-Guerraoui}  
we focus on locks closely related to our design. 

Simple test-and-set or polite test-and-test-and-set \cite{ScottB24} locks 
are compact and exhibit excellent latency for uncontended operations, but fail 
to scale and may allow unfairness and even indefinite starvation.  Ticket Locks are compact
and FIFO, and also have excellent latency for uncontended operations, but they also 
fail to scale because of global spinning, although some variations attempt
to overcome this obstacle, at the cost of increased space
\cite{EuroPar19-TWA,Ticket-AWN}.  
For instance, Anderson’s array-based queueing lock \cite{dc03-Anderson,tpds90-Anderson} 
is based on Ticket Locks but provides local spinning.  It employs a waiting 
array for each lock instance, sized to ensure there is at least one array element 
for each potentially waiting thread, yielding a potentially large footprint. 
The maximum number of participating threads must be known in advance when
initializing the lock.  TWA\cite{EuroPar19-TWA} is a variation on ticket locks
that reduces the incidence of global spinning.

Queue-based locks such as MCS or CLH are FIFO and provide local spinning and
are thus more scalable.  
MCS is used in the linux kernel for the low-level ``qspinlock'' construct
\cite{linux-locks,Long13,LWN2014}.  Modern extensions of MCS edit the queue
order to make the lock \emph{NUMA-Aware}\cite{EuroSys19-CNA}. 
MCS readily allows editing and re-ordering of the queue of waiting threads,
\cite{markatos,eurosys17-dice,EuroSys19-CNA} whereas editing the
chain is more difficult under CLH, HemLock and \recipro{}, where
there are no explicit linked lists.

CLH is extremely simple, has excellent RMR complexity, and requires just an 
single atomic exchange operation in the \Acquire{} operation and no atomic 
read-modify-write instructions in \Release{}. 
Unfortunately the waiting elements migrate between threads, which may be
inimical to performance in NUMA environments.  CLH locks also require 
explicit constructors and destructors, which may be inconvenient
\footnote{ 
Many lock implementations require just trivial constructors to set the lock fields to 0 or some constant,
and trivial destructors, which do nothing.  The Linux kernel \texttt{spinlock\_t}/\texttt{qspinlock\_t}
interface provides a constructor, but does not even expose a destructor.
Similarly, C++ \texttt{std::mutex} is allowed to be \emph{trivially destructible}, meaning
storage occupied by trivially destructible objects may be reused without calling the destructor.  
Under both GCC g++ version 13 and Clang++ Version 18, \texttt{is\_trivially\_destructible} reports \texttt{True}
for \texttt{std::mutex}, and as such, destructors do not run.  This situation could be rectified by
recompiling, but we would also need to recompile all libraries transitively depended on by
the application, including those deployed in shared libraries in binary-only form on the system.}. 
Our specific implementation uses a variation on Scott's \cite{ScottB24} Figure 4.14, which
converts the CLH lock to be \emph{context-free}\cite{ppopp16-wang}, adhering to a simple 
programming interface that passes just the address of the lock, albeit at the cost of
adding an extra field to the lock body to convey the address of the
head waiting element to \Release{}.  


The K42 \cite{K42,ScottB24} variation of MCS can recover the queue element before returning from \Acquire{}
whereas classic MCS recovers the queue element in \Release{}.  
That is, under K42, a queue element is needed only while waiting but not while the lock is held,
and as such, queue elements can always be allocated on stack, if desired. 
While appealing, the paths are much more complex and touch more cache lines than the classic
version, impacting performance.  In addition, neither the doorway nor the \Release{} path operate
in constant time.

HemLock combines aspects of both CLH and MCS to form a lock
that has very simple waiting node lifecycle constraints, is \emph{context-free} but
still scales well in common usage scenarios.  
HemLock does not provide constant remote memory reference (RMR) complexity 
\cite{opodis17-dvir} in scenarios where a thread holds multiple contended locks. 
In this situation the single node suffers from \emph{multi-waiting}\cite{spaa21-Dice}.  
Similar to MCS, HemLock lacks a constant-time unlock operation, whereas
the unlock operator for CLH and Tickets is constant-time.  Unlike MCS, HemLock requires
active synchronous back-and-forth communication in the unlock path between the outgoing
thread and its successor to protect the lifecycle of the waiting element. 
We note, however, that HemLock remains constant-time in the \Release{} operator to the
point where ownership is conveyed to the successor.  
HemLock uses \emph{address-based} transfer of ownership, writing the address
of the lock instead of a boolean, differentiating it from MCS and CLH. 
\recipro{}, like HemLock, requires just a singleton per-thread waiting element allocated
in thread-local storage.  


The closest related work to \recipro{} that we know of is Chen's mutual exclusion 
algorithm\cite{icpads05-chen,tpds09-chen},
where arriving threads use an atomic \texttt{exchange} on arrival to either acquire
the lock or join a LIFO stack of recently arrived waiting threads.  
A new stack is detached (or \emph{closed} in their terminology) when the current stack is exhausted.
Their progress and bounded bypass properties are the same as found in \recipro{}.   
Their approach uses global spinning, however, and requires at least 2 shared global variables -- above
and beyond context passed from \Acquire{} to \Release{} in the form of the immediate successor.  
In addition, each \Release{} mutates at least one global shared variable, increasing coherence traffic. 


We opted to exclude NUMA-aware locks such as Cohort Locks \cite{topc15-dice,PPoPP12-Dice} and 
Compact NUMA-Aware Locks (CNA) \cite{EuroSys19-CNA} from consideration.
We excluded Fissile Locks and GCR\cite{EuroPar19-GCR} as they have 
lax time-based anti-starvation mechanism.
Fissile, specifically, depends on the owner of the \emph{inner lock} to make progress
to monitor for starvation.  

We also excluded MCSH\cite{MCSH} which is a recent variation of MCS
that uses on-stack allocation of queue nodes and hence supports a standard locking interface.  
Like MCS, the lock is not innately context-free, and additional information needs to
be passed from \Acquire{} to \Release{} via extra fields in the lock body.  
In our experiments, the performance of MCSH is typically on par with MCS proper.

We further restrict our comparison to locks that use \emph{direct succession} and 
hand off ownership directly from the owner to a specific successor and
do not admit \emph{barging} or \emph{pouncing} where the lock is released even though
waiting threads exist, which allow newly arriving threads to sieze the lock.  


\Invisible{Exclude locks that use \emph{barging}} 

\Invisible{We excluded MCSH\cite{MCSH}, which is essentially a simplified Fissile lock that uses MCS instead 
of MCS-based NUMA-aware CNA for the inner lock, removes the fast-path \emph{barging} 
attempt on arrival, has \emph{patience} set to 0, yielding FIFO admission, and uses 
Fissile's ''partial MCS release'' optimization to defer the final step in releasing the 
inner lock until the \Release{} operation.}

\newcommand{\NOTEA}{\hyperlink{HemLock}{\blu{Note-1}}}

\setlength{\thickmuskip}{0mu}
\setlength{\thinmuskip}{0mu}
\setlength{\medmuskip}{0mu}

\begin{table*} [ht!]
\footnotesize
\centering
\begin{tabular}{lllllll}
\toprule

\multicolumn{1}{l}{} &
\multicolumn{6}{c}{Lock Algorithm}      \\
\cmidrule(lr){2-7}

\multicolumn{1}{l}{Property} &
\multicolumn{1}{l}{\rot{MCS}} &
\multicolumn{1}{l}{\rot{CLH}} & 
\multicolumn{1}{l}{\rot{HemLock}} & 
\multicolumn{1}{l}{\rot{Ticket}} & 
\multicolumn{1}{l}{\rot{TWA}} &
\multicolumn{1}{l}{\rot{Reciprocating}} \\
\midrule
Spinning                             & Local     & Local     & \red{Semi}  & \red{Global}   & \red{Semi}       & Local    \\
Constant-time \Release{}             & \red{No}  & Yes       & \red{No}    & Yes            & Yes              & Yes      \\
Context-free                         & \red{No}  & \red{No}  & Yes         & Yes            & Yes              & \red{No} \\ 
FIFO                                 & Yes       & Yes       & Yes         & Yes            & Yes              & \red{No} \\
Path Complexity -- \Acquire{}        & 29        & 19        & 22          & 12             & 57               & 42       \\
Path Complexity -- \Release{}        & 17        & 4         & 21          & 5              & 11               & 20       \\ 
On-Stack                             & No        & No        & No          & N/A            & N/A              & Possible \\ 
Nodes Circulate                      & No        & \red{Yes} & No          & N/A            & N/A              & No       \\
Explicit CTOR/DTOR Required          & No        & \red{Yes} & No          & No             & No               & No       \\
Maximum Remote Misses per episode    & 4         & 4         & \blu{4}     & \red{$T$}      & \red{$T\cdot{}2$} & 2        \\
Invalidations per episode            & 6         & 5         &  5          & \red{10($T$)}  & \red{8.5($T$)}    & 4        \\
Space Complexity                     & $(S\cdot{}L)+(E\cdot{}A)$ & $(S\cdot{}L) + (E\cdot{}(L+T))$ & $L + (E\cdot{}T)$ & $S\cdot{}L$  & $(S\cdot{}L) + 4096$  & $(S\cdot{}L) + (E\cdot{}T)$ \\

\midrule[\heavyrulewidth]
\bottomrule
\end{tabular}%
\caption{Comparison of lock algorithm properties}\label{Table:Compare}
\end{table*}

In Table-\ref{Table:Compare} we compare the attributes of various local algorithms.  
\red{Red}-colored cells indicate potential undesirable properties.  
Note that all the locks provide a constant-time doorway phase.
In the following we explain the meaning of each property.   

\textbf{Spinning}: reflects whether the lock utilizes \emph{local}, 
\emph{global} or \emph{semi-local}\cite{spaa21-Dice} spinning.  
\textbf{Constant-time Unlock}: indicates if the lock \Release{} operation is bounded.  
As noted above, HemLock is constant-time up to the point where the lock is either released or transferred to a successor, 
but the \Release{} operator, for reasons of memory safety, then waits for the successor to 
acknowledge transfer of ownership before the memory underlying the queue element can be potentially reused.
Specifically, in HemLock an uncontended \Release{} operation is constant-time and a contended \Release{} is 
constant-time up to and including the point where ownership is conveyed to the successor.  
\textbf{FIFO}: indicates the lock provides strict FIFO admission.  
\textbf{Context-free}: indicates additional information does not need to be transferred from the lock
operator to the corresponding \Release{} operation.
\textbf{Path Complexity - \Acquire{}} and \textbf{Path Complexity - \Release{}} : 
we measured the size of the \Acquire{} and \Release{} methods in units of 
platform independent LLVM intermediate representation (IR) instructions, as emitted by \texttt{clang++-18},
which serves as a simple measure for path complexity.  We note that much of the complexity found in TWA manifests through the 
use of the hash function which maps lock address and ticket value pairs to slots in the waiting array. 
\textbf{On-Stack}: indicates the queue elements, if any, may be allocated on-stack.  This also
implies the nodes do not migrate and have a tenure constrained to the duration of the locking episode. 
\textbf{Nodes Circulate}: queue elements migrate between threads.  This often implies the need for
an explicit queue element lifecycle management system and precludes convenient on-stack allocation
of queue elements.  Migration may also be unfriendly to performance in NUMA environments.  
\textbf{Explicit CTOR/DTOR Required}: indicates the lock requires non-trivial constructors or destructors.  
CLH, for instance, requires destructors to run to release the wait elements referenced in the lock, to
avoid memory leaks.   
\textbf{Maximum Remote Misses per episode}: is the worst-case maximum number of misses to \emph{remote} memory
incurred, under simple sustained contention, by a matching \Acquire{}-\Release{} pair.  Misses to \emph{remote} memory --
memory \emph{homed} on a different NUMA node --
may be more expensive than local misses on various platforms, such as modern Intel systems that use 
the UPI coherence fabric \cite{UPI}, where miss requests are first adjudicated by the \emph{home node} of the cache line. 
Algorithms where nodes circulate appear more vulnerable to accumulating such remote misses.  
For \blu{HemLock}, we assume simple contention with no \emph{multi-waiting}.  
We derived the ``Maximum Remote Misses per episode'' value via static inspection of the code, 
and determining which of the accesses that might cause coherence misses might also be to locations 
homed on remote nodes. 
\textbf{Invalidations per episode}:  is the number of coherence misses, under sustained contention,
experienced by an \Acquire{}-\Release{} episode in a given thread.   
We empirically approximate that number as follows.  
Using an ARMv8 system (described below), we modifed the MutexBench microbenchmark to have a degenerate critical section that advanced
only a \emph{local} random number generator and to pass any context from \Acquire{} to \Release{} via 
thread-local storage, in order to reduce mutation of shared memory.  Absent coherence misses, an \Acquire{}-\Release{} episode,
including the critical section, manages to remain completely resident in the private L1 data cache. As such, any misses
are coherence misses.  
This technique yields a useful empirical metric on how much coherence traffic various lock algorithms generate.
We used the ARM \texttt{l2d\_cache\_inval} hardware performance counter, which 
tallies L2 cache invalidation events, as simple proxy for coherence traffic -- both latency and bandwith.  
We report the number of \texttt{l2d\_cache\_inval} events per episode. As can be seen, Reciprocating Locks
is parsimonious and incurs just 4 invalidations per episode, while CLH requires 5
\footnote{We observed similar ratios with performance counters that reported the number of ``snoop'' messages,
and on Intel by counting \texttt{OFFCORE} requests.
We also observed in passing, while examining data from hardware performance counters, that CLH and MCS suffered 
from more stalls (both event counts and duration) from misses than did Reciprocating locks.  
CLH in particular executes a dependent load in the critical path in the arrival doorway on the address returned 
from the atomic \texttt{exchange}. 
The address to be loaded from is not known until after the \texttt{exchange} returns,
denying the processor the ability to speculate or execute out-of-order.}.  
Ticket locks require $T$ (10 in our example) invalidations per episode, where $T$ is the number of participating threads, 
given the global spinning.  The number of misses incurred by CLH, MCS, HemLock and Reciprocating Locks is
constant and not a function of the number of threads. 
These empirically-derived results align closely with a static analysis of the code and the expected number of 
coherence misses in the \Acquire{} and \Release{} paths.  
\textbf{Space Complexity}: reflects the space complexity of the lock algorithm where $T$ is the number of active threads,
$L$ is the number of extant locks, $A$ is the number of locks currently held plus the number
of threads currently waiting on locks, $E$ is the size of the waiting element. 
$S$ is the size of a lock instance.  When a lock algorithm
requires context to be passed from \Acquire{} to the corresponding \Release{}, we set $S$ to $2$
to indicate that our implementation allocated an extra word in the lock body to pass such information. 
Implementations that require context can avoid that space requirement in the lock body if they 
opt to pass context information by other means, such as via thread-local storage, in which case $S$ would be $1$.  
HemLock is context-free, so the per-lock space usage is just $1$, while
Ticket Locks and TWA require 2 words per lock.  
TWA requires an additional 4096 words for the global waiting table that is shared over all thread and lock instances. 
For MCS and CLH we assume that the implementation stores the head of the chain -- reflecting the current 
owner --  in an additional field in the lock body, and thus the lock consists of 
\texttt{head} and \texttt{tail} fields, requiring 2 words in total.

\Invisible{semi-local; commonly, frequently, often, usually mostly ::: fere- ; pene- ; quotide ; semi-; cotidie ; vulgo-; plerumque ; saepe- }

\section{Empirical Performance Results}

Unless otherwise noted, all data was collected on an Oracle X5-2 system
system having 2 sockets, each populated with
an Intel Xeon E5-2699 v3 CPU running at 2.30GHz.  Each socket has 18 cores, and each core is 2-way
hyperthreaded, yielding 72 logical CPUs in total.  The system was running Ubuntu 20.04 with a stock
Linux version 5.4 kernel, and all software was compiled using the provided GCC version 9.3 toolchain
at optimization level ``-O3''.
64-bit C or C++ code was used for all experiments.
Factory-provided system defaults were used in all cases, and Turbo mode \cite{turbo} was left enabled.
In all cases default free-range unbound threads were used, with no pinning of threads to processors. 
At above 18 ready threads, the kernel starts to place additional threads onto the second socket,
and NUMA effects come into play. 

We implemented all user-mode locks within LD\_PRELOAD interposition
libraries that expose the standard POSIX \texttt{pthread\_\allowbreak{}mutex\_t} programming interface
using the framework from ~\cite{topc15-dice}.
This allows us to change lock implementations by varying the LD\_PRELOAD environment variable and
without modifying the application code that uses locks.
The C++ \texttt{std::mutex} construct is implemented directly via \texttt{pthread\_\allowbreak{}mutex} primitives,
so interposition works for both C and C++ code.
All lock busy-wait loops on Intel used the \texttt{PAUSE} instruction and \texttt{YIELD} on ARMv8.  
To reduce false sharing, all waiting elements were aligned and sequestered at 
128-byte boundaries, and, for the ``MutexBench'' benchmarks, below, the lock instances
where similarly sequestered. 
For the CLH, MCS, and \recipro{} implementations used in the benchmark section of this paper, 
we elected to pass any additional context information via extra fields in the lock instance. 


\subsection{MutexBench benchmark} 
\label{Section:MutexBench} 

\renewcommand{\topfraction}{0.85}
\renewcommand{\bottomfraction}{0.85}
\renewcommand{\textfraction}{0.15}
\renewcommand{\floatpagefraction}{0.8}
\renewcommand{\textfraction}{0.1}
\setlength{\floatsep}{5pt plus 2pt minus 2pt}
\setlength{\textfloatsep}{5pt plus 2pt minus 2pt}
\setlength{\intextsep}{5pt plus 2pt minus 2pt}
\setlength{\abovecaptionskip}{1pt} 
\setlength{\belowcaptionskip}{1pt} 

\begin{figure*}[t!]
\subfloat[Maximum Contention Intel]{%
    \includegraphics[width=7cm]{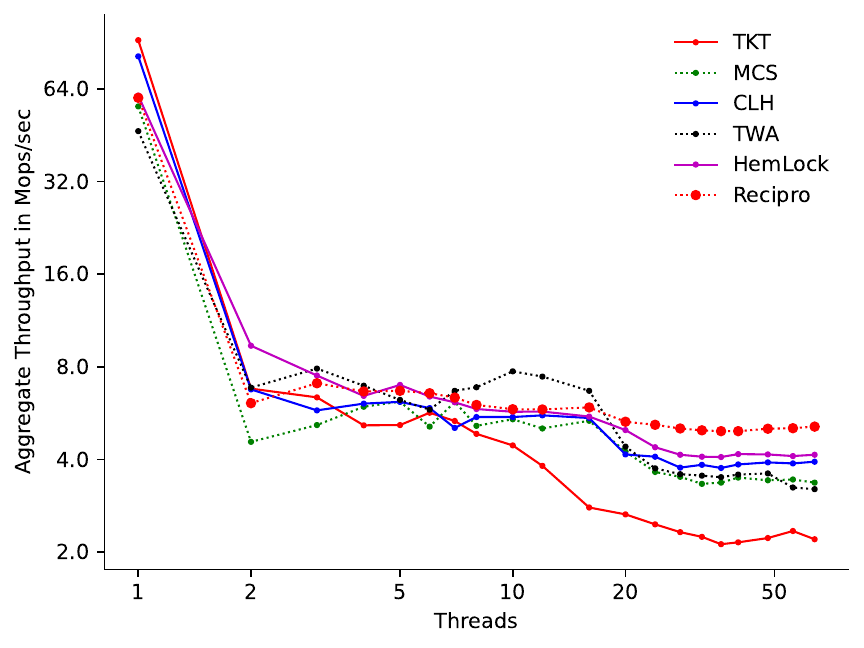}
    \label{Figure:MaximumContention}                                                             
}\hfill
\subfloat[Moderate Contention Intel]{%
    \includegraphics[width=7cm]{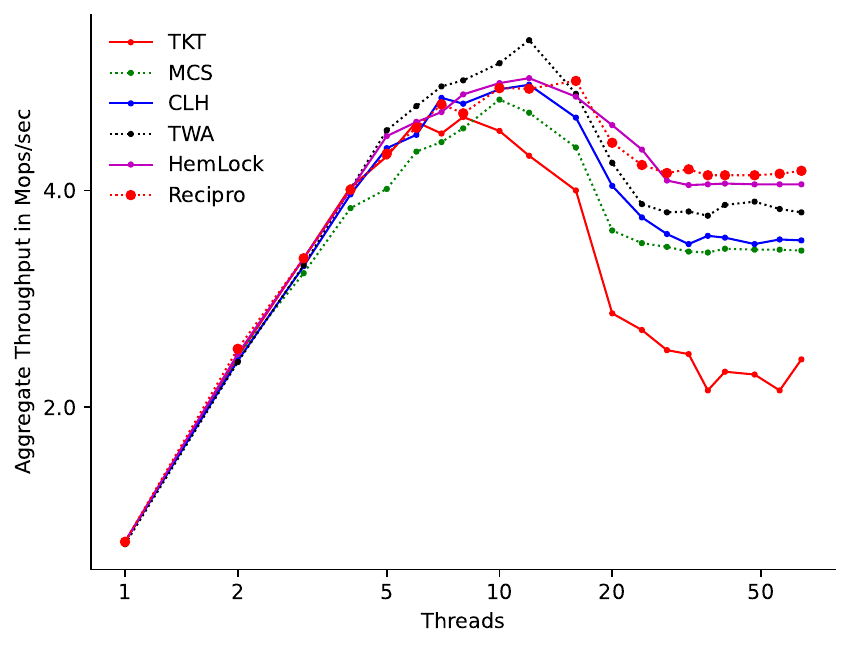} 
    \label{Figure:ModerateContention}                                                             
}\\
\subfloat[Maximum Contention ARMv8]{%
    \includegraphics[width=7cm]{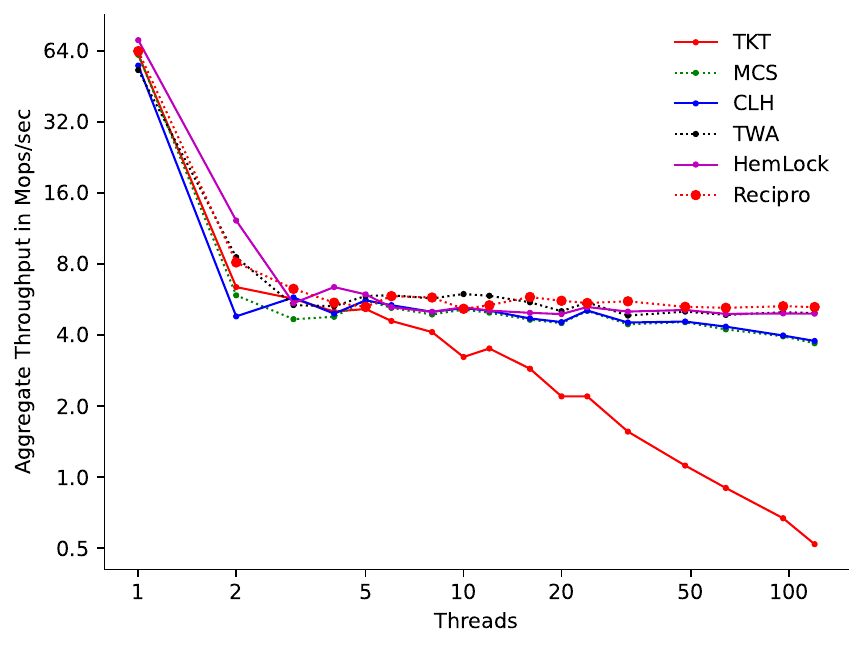}
    \label{Figure:MaximumContention-armv8}                                                             
}\hfill
\subfloat[Moderate Contention ARMv8]{%
    \includegraphics[width=7cm]{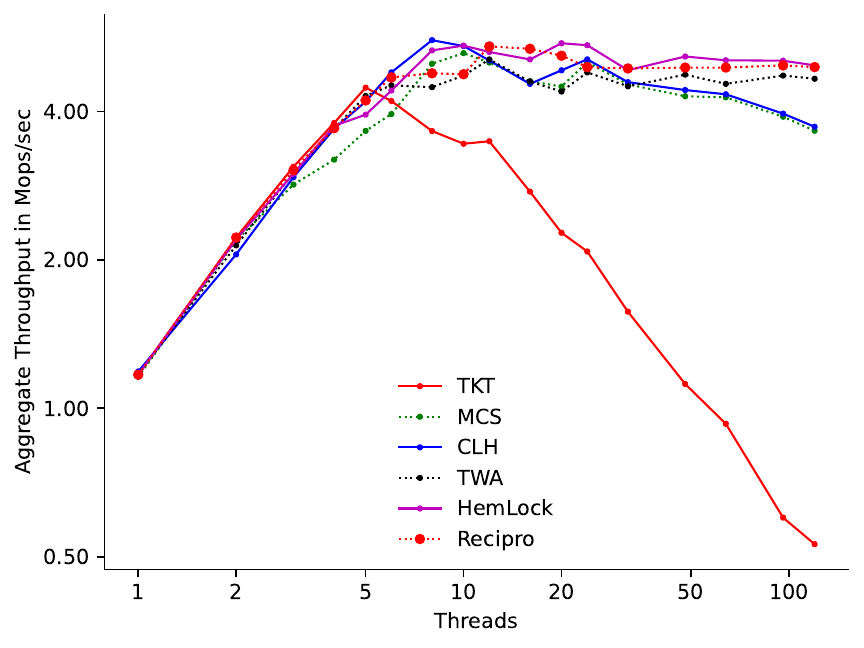} 
    \label{Figure:ModerateContention-armv8}                                                             
}
\caption{MutexBench}
\label{fig:MutexBench}
\end{figure*}


The MutexBench benchmark spawns $T$ concurrent threads. Each thread loops as follows:
acquire a central lock L; execute a critical section; release L; execute
a non-critical section. At the end of a 10 second measurement interval the benchmark
reports the total number of aggregate iterations completed by all the threads.
We report the median of 7 independent runs in Figure-\ref{Figure:MaximumContention}
where the critical section advances a
shared global \texttt{std::mt19937} \emph{Mersenne Twister} pseudo-random number generator (PRNG)
one step, and the non-critical section is empty, 
subjecting the lock to extreme contention.  (At just one thread, this configuration also constitutes
a useful benchmark for uncontended latency).  
The $X$-axis reflects the number of concurrently executing threads contending for the
lock, and the $Y$-axis reports aggregate throughput -- 
the tally of all loops executed by all the threads in the measurement interval. 
For clarity and to convey the maximum amount of information to allow a comparison of the algorithms,
the $Y$-axis is offset to the minimum score and the $X$-axis is logarithmic.

We ran the benchmark under the following lock algorithms: 
\textbf{MCS} is classic MCS.  
To avoid memory allocation during the measurement interval,
the MCS implementation uses a thread-local stack of free queue elements. 
\textbf{CLH} is CLH based on Scott's \emph{CLH variant with a standard interface} Figure-4.14 of \cite{ScottB24}; 
For the MCS and CLH locks, our implementation stores the current head of 
the queue -- the owner -- in a field adjacent to the tail, so the lock body 
size was 2 words.  
CLH presents something of a challenge when used under the \texttt{pthread\_mutex} interface. 
First, pthreads allows the programmer to use trivial initializers -- setting the mutex body to all 0 -- 
and avoid calling \texttt{pthread\_mutex\_init}.  To compensate, we modified \texttt{pthread\_mutex\_lock}
to populate such an uninitialized lock with the CLH ``dummy node'' lazily, on-demand, on the first lock operation.
In \texttt{pthread\_mutex\_destroy} we free the node, if populated, but many applications also do not 
call \texttt{pthread\_mutex\_destroy}, which constitutes a memory leak.  As such, to allow CLH to be 
included, we avoided applications
that create and then abandon large numbers of locks in a dynamic fashion.
\textbf{Ticket} is a classic Ticket Lock;
\textbf{HemLock} is the HemLock algorithm, with the CTR (coherence traffic reduction) optimization.


As we are implementing a general purpose pthreads locking interface, a 
thread can hold multiple locks at one time.  Using MCS as an example, lets say thread 
$T1$ currently holds locks $L1$,$L2$,$L3$ and $L4$.  We'll assume no contention.
$T1$ will have deposited MCS queue nodes into each of those locks.  MCS nodes 
can not be reclaimed until the corresponding unlock operation.  Our implementation 
could \texttt{malloc} and \texttt{free} nodes as necessary -- allocating in the 
lock operator and freeing in unlock -- but to avoid malloc and its locks, we 
instead use a thread-local stack of free queue nodes.  In the lock operator, 
we first try to allocate from that free list, and then fall back to \texttt{malloc}
only as necessary.  In unlock, we return nodes to that free list.  This approach 
reduces \texttt{malloc-free} traffic and the incumbent scalability concerns.   
We currently don't bother to trim the thread-local stack of free elements.  
So, if thread $T1$ currently holds no locks, the free stack will contain $N$ elements 
where $N$ is the maximum number of locks concurrently held by $T1$.  We 
reclaim the elements from the stack when $T1$ exits.  A stack is convenient for locality.  

In Figure-\ref{Figure:MaximumContention} we make the following observations regarding
operation at maximal contention with an empty critical section:
\begin{itemize}[align=left,leftmargin=2em,labelwidth=0.8em]
\item At 1 thread the benchmark measures the latency of uncontended \Acquire{} and \Release{} operations.
Ticket Locks are the fastest, followed closely by HemLock, \recipro{}, CLH and MCS. 
\item As we increase the number of threads, Ticket Locks initially do well but then fade,
exhibiting a precipitous drop in performance.  TWA is the clear leader in the ``middle'' area
of the graph, between 4 and 16 threads.  
\item Broadly, at higher thread counts, HemLock performs slightly better than or the same as CLH or MCS,
while \recipro{} provides the best throughput. 
\end{itemize} 

We note in passing that care must be taken when \emph{negative} or \emph{retrograde}
scaling occurs and aggregate performance degrades as we increase threads.
As a thought experiment, if a hypothetical lock implementation were to introduce
additional synthetic delays outside the critical path, aggregate performance might increase as
the delay throttles the arrival rate and concurrency over the contended lock \cite{SIF}. 
As such, evaluating just the maximal contention case in isolation is insufficient.

In Figure-\ref{Figure:ModerateContention} we pass arguments to MutexBench that configure it 
to implement a delay in the non-critical section.  Each thread has a private 
\texttt{std::mt19937} pseudo-random number generator.
The non-critical section generates, using the private PRNG, a uniform random number in the range $[0-250)$ and then
advances its private PRNG that many steps.  (We take care to make sure this 
operation can not be optimized away by consuming the PRNG value at the end of the thread's run).  
As above, the critical section advances the shared PRNG one step.  
As can be seen, we enjoy positive scalability up to about 5 threads.  At higher thread counts,
\recipro{} manages to exhibit the least reduction in scalability.

 


In Figure-\ref{Figure:MaximumContention-armv8} and Figure-\ref{Figure:ModerateContention-armv8},
we show the results of MutexBench on an ARMv8 (aarch64) system, showing that the relative performance of
the various algorithms remains portable over disparate architectures.  The ARMv8 system was an Ampere Altra Max
NeoVerse-N1 with 128 processors on a single socket and was running Ubuntu 24.04.  
We compiled all code using the \texttt{-mno-outline-atomics} \texttt{-march=\allowbreak{}armv8.2-a+lse} flags
in order to allow direct use of modern atomic \texttt{exchange}, \texttt{CAS} and \texttt{fetch-and-add} instructions instead
of the legacy LL-SC (load-locked store-conditional) forms thereof.

\subsection{std::atomic}


\begin{figure*}[ht!]
\subfloat[exchange()]{%
    \includegraphics[width=7cm]{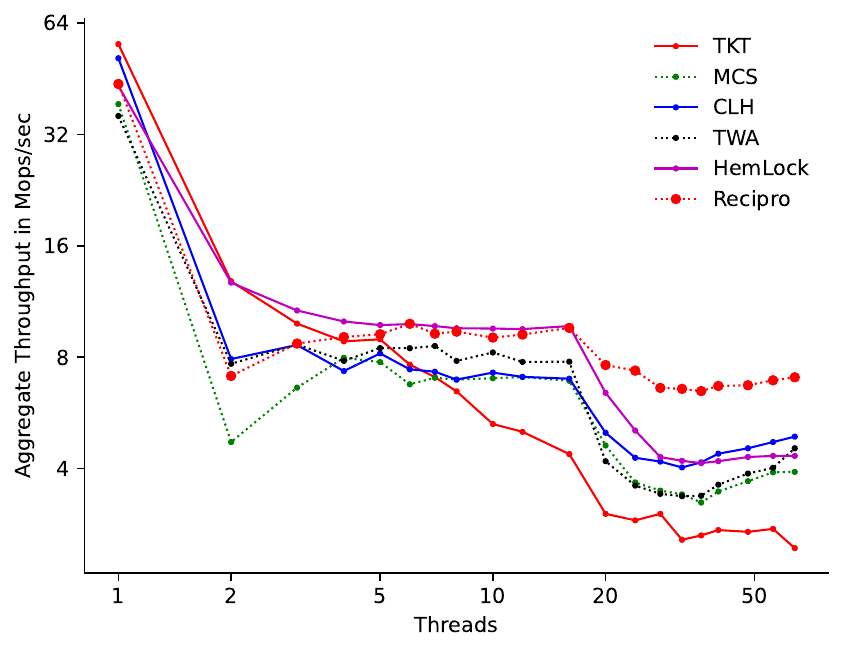} 
    \label{Figure:atomicstruct}                                                             
}\hfill
\subfloat[compare\_exchange\_strong()]{%
    \includegraphics[width=7cm]{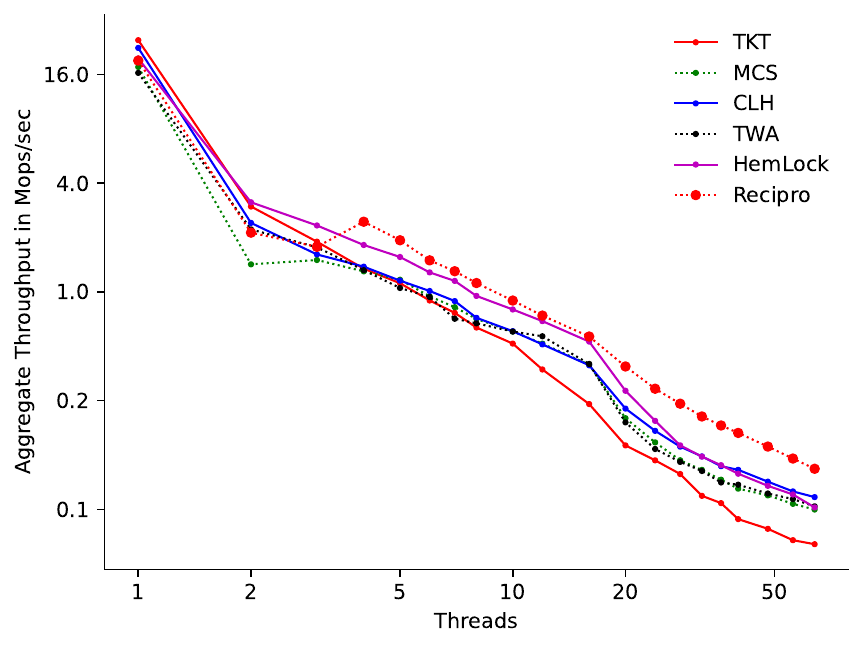} 
    \label{Figure:atomiccas}                                                             
}
\caption{C++ std::atomic<struct>}
\label{fig:atomic}
\end{figure*}

In Figure-\ref{Figure:atomicstruct}, our C++ atomic exchange benchmark defines a simple structure 
type \texttt{S} that contains 5 32-bit integers.  
Each thread declares a local \texttt{S} instance, and we also declare a shared
global \texttt{std::atomic<S> A \{\}}.    
The C++ compiler and runtime implement \texttt{std::atomic} for such objects by hashing
the address of the instance into an array of mutexes, and acquiring those as needed to implement
the desired atomic action.  Each thread loops for 10 seconds, using \texttt{std::atomic<S>::exchange} to
swap its local copy with the shared global, and we then report aggregate throughput rate, in terms
of completed exchange operations, at the end of the measurement interval.  
The figure depicts the median score of 7 runs and the overall ranking of the locks is similar
to that observed under MutexBench.


Figure-\ref{Figure:atomiccas} is similar, but instead of \texttt{std::atomic:exchange}, we
use \texttt{std::atomic:load} to fetch the shared global, make a local copy, increment
the 1st of the five constituent integer fields in that copy, and then call \texttt{std::atomic:compare\_\allowbreak{}exchange\_\allowbreak{}strong}
(CAS) in an inner loop to try replace the global copy with the updated local version.  The \texttt{load} operator acquires and releases
all necessary covering locks, as does the \texttt{compare\_\allowbreak{}exchange\_\allowbreak{}strong} operator.
If the CAS fails, we update our fresh local copy (returned from CAS) and retry. 
At the end of the 10 second measurement interval we report total number of successful CAS operations.
The graph reports the median score of 7 runs.
For both benchmarks, the relative ranks of the various lock algorithms remains similar
to what was observed with MutexBench.

\subsection{LevelDB}  

\begin{figure}[h!]                                                                    
\includegraphics[width=7cm]{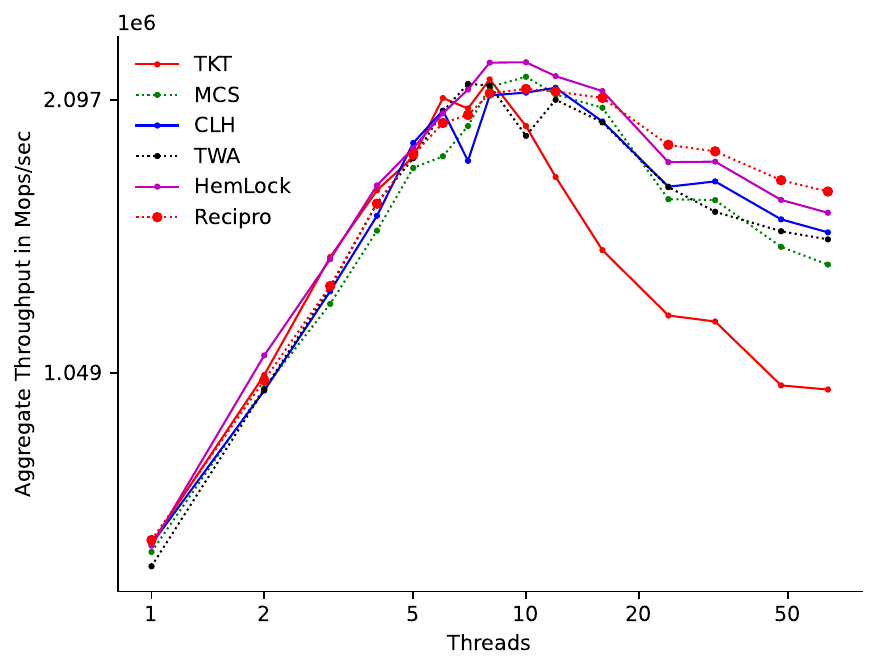} 
\caption{LevelDB : readrandom}                                                              
\label{Figure:LevelDB}                                                             
\end{figure}          

In Figure-\ref{Figure:LevelDB}  we used the ``readrandom'' benchmark in LevelDB version 1.20
database\footnote{\url{leveldb.org}} varying the number of threads and reporting throughput
from the median of 5 runs of 50 seconds each.
Each thread loops, generating random keys and then tries to read the associated value from
the database.
We first populated a database\footnote{db\_bench \---\---threads=1
\textendash{}\textendash{}benchmarks=fillseq \---\---db=/tmp/db/}
and then collected data\footnote{db\_bench \---\---threads=\emph{threads}
\---\---benchmarks=readrandom \\ \---\---use\_existing\_db=1
\---\---db=/tmp/db/ \---\---duration=50}.
We made a slight modification to the \texttt{db\_bench} benchmarking
harness to allow runs with a fixed duration that reported aggregate throughput.
LevelDB uses coarse-grained locking, protecting the database with a single central mutex:
\texttt{DBImpl::Mutex}.  Profiling indicates contention on that lock via \texttt{leveldb::DBImpl::Get()}.
The results in Figure-\ref{Figure:LevelDB} largely echo those in Figure-\ref{Figure:ModerateContention}.

\section{Discussion} 




\Texticle{}Constant-time arrival doorway and \Release{} paths, as
a consequence of having desirable progress properties, may yield more predictable performance.  
Furthermore, such designs are more inherently amenable to ``polite'' waiting,
where a thread voluntarily deschedules itself and explicitly waits to be notified of transfer of ownership, by
means of \emph{thread identity-based waiting} primitives, such as park-unpark\cite{arxiv-Malthusian} or
\emph{address-based waiting}, such as the linux \emph{futex}\cite{futex} primitive, which 
underlies the pthreads synchronization services.  Use of such primitives allows the implementation
to avoid toxic waiting policies, such as unbounded busy-waiting;  periodic polling via timed sleep operations; 
or operating-system advisory \texttt{yield} calls.  Specifically, having constant-time paths means
contending threads wait on only one condition -- transfer of ownership -- and have one waiting phase.  In contrast, locks such
as MCS or HemLock may need to wait on additional conditions.  


\Texticle{}Lifecycle management : In classic MCS the address of the owner's ``queue node'' must
be passed from an acquire operation to the corresponding release.  Relatedly,
that queue node must remain extant during the entire acquire-release period,
and can not be use for other locking operations.
In \recipro{}, the element needs to remain extant only for the
duration of the acquire operation, affording the implementer more latitude
as to how and where the element is allocated.  A singleton element allocated in thread-
local storage suffices.

\Texticle{}Passing of context from acquire to release :
Like MCS and CLH, \recipro{} requires context to be passed from
an acquire operation to the corresponding release.  
Broadly, any context is undesirable and to the extent possible, we might prefer \emph{context-free} algorithms.
In practice, it is common to keep required context in extra fields in the lock body (which
can induce increased coherence traffic and increase the size of the lock)
or in thread-local storage constructs that track which locks are held by a
thread and can also be used to convey context.  In managed runtime environments
such as the HotSpot Java Virtual Machine, with balanced locking under the ``synchronized''
construct, such context information can be kept in the stack frame.  In C++,
context can be kept in \texttt{std::lock\_guard} or \texttt{std::scoped\_lock} instance,
which is typically allocated on stack.
Such constructions are an example of the the \emph{Resource Acquisition is Initialization(RAII)}\cite{RAII} idiom.
And in general, we observe that more flexible locking interfaces that allow context to be easily passed,
outside of the lock body, will allow a broader range of lock algorithms to be deployed.  

\Invisible{Lessons for locking API designers regarding the interplay between algorithms and interfaces}

\Texticle{}Waiting phases: in a given \Acquire{}-\Release{} episode, a thread
will require at most one waiting phase under \recipro{}, while MCS and HemLock may require up to two phases. 
This makes \recipro{} more amenable to waiting via \emph{futex} (address-based) or \emph{park-unpark}
(identity-based) constructs which deschedule waiting threads, making them ineligible to be 
dispatched on CPUs by the scheduler until they are explicitly notified and again made ready (runnable).  
Hapax locks \cite{PPoPP26-dice,dice2026hapaxlocksvaluebased} also require just one waiting phase.  

\Texticle{}Handoff costs: Under sustained steady-state contention, CLH and MCS \Release{} operators can 
perform handover without any need to access the central shared lock body, via
direct thread-to-thread communication that does not involve accesses to the shared lock body. 
This desireable property acts to reduce coherence miss costs and improve scalability.  
\recipro{}, however, needs to occasionally consult the central lock \texttt{arrivals} pointer 
when the entry segment is found to be exhausted and must be reprovisioned.  
Compared to CLH and MCS, this can increase RMR complexity and generate additional coherence traffic.
Interestingly, as we increase the number of contending threads, reciprocating locks will tend 
to enjoy longer segments and thus need to access the central arrival word less often. 
In addition, the conditional branch found in the release path of reciprocating locks -- 
which tests whether the successor is \emph{nullptr} -- will tend to be
less well-predicted when the segments are shorter.

\Texticle{} As to why \recipro{} sometimes exceeds the performance of CLH, we make
the following observations : 

\vspace{-4pt} 

\begin{enumerate}[label=(\Alph*),labelindent=0pt,wide,labelwidth=!]
\item on NUMA systems, the waiting element in \recipro{}
is almost surely \emph{homed} on same NUMA node as the associated thread, providing a performance
benefit if home-based snooping is employed.  Under CLH, the elements circulate.
We note, however, that \recipro{} still tends to outperform CLH on non-NUMA systems,
although the difference is lessened.

\item \recipro{} incurs less indirection and suffers less from stalls
arising from address-dependent data loads.  Under CLH, the address to be waited upon is not known
until after the atomic exchange completes -- and the coherence miss from the exchange is satisfied --
whereas in \recipro{} the address to be waited upon under contention is known and
fixed for the thread.

Similarly, consider the step in CLH before the atomic exchange where a
thread clears the flag in the element about to be installed via an atomic exchange.
Assuming the implementation of CLH found in Figure 4.14  of
\cite{ScottB24}, thread-local storage contains a pointer
to the element, which in turn holds the flag.  We find a load from thread-local storage
to fetch the address of the element, followed by a dependent store into the element.
Under \recipro{}, thread-local storage simply contains the flag proper, embedded
in the \texttt{WaitingElement}, reducing indirection.  We thus require just a store to
prepare the element.
Broadly, CLH requires more indirection
because elements circulate.  The address of the element to be installed via exchange,
and the address of the element to be waited upon will, for a given thread, vary
between iterations.  In \recipro{}, those addresses are fixed for the thread.

\item If we tally the minimum number of coherence misses
in an idealized contended acquire-release episode, for CLH we have 5 : the store to
prepare the element before the exchange; the exchange; the 1st load in the waiting phase
after the exchange; the last load in the waiting phase when ownership is passed;  
and finally, the store in the release operation that conveys ownership.
The store to prepare the element, prior to the exchange, causes a coherence miss as the last 
store to the body of that element was from a different thread, before the element migrated to
the thread in question. (This miss occurs outside and before the critical section, but is
still undesirable as it causes a stall and consumes interconnect bandwidth).
For \recipro{}, in Listing-\ref{Listing:WFvAB-lambda}, we incur a simple
\emph{shared$\rightarrow$modified} coherence upgrade at Line-17 (the line in already 
in \emph{shared} state, because of the previous \Acquire{} operation), a miss at Line-20 to exchange the
address of the element, a miss at Line-30 when the busy-wait loop condition is ultimately satisfied
and ownership is conveyed, and, in the \Release{} operation, a miss at Line-58 when
transferring ownership to the successor, for a total of 4 coherence misses.  
The misses at Line-17 and Line-30 are to memory expected to be \emph{homed} on
the same NUMA node as the thread.

\end{enumerate} 



\section{Palindromic Admission Schedules} 

\Invisible{Relegate; demote; defer; aside } 
\Invisible{epi-cycles, epi-palindromes} 


\recipro{} may allow a \emph{palindromic} admission schedule which, under the right
circumstances, can persist for long periods.  

\subsection{Example Scenario} 
Table-\ref{Table:Palindromic} illustrates the phenomena with a simple scenario.
Threads $A$ $B$ $C$ $D$ and $E$ all complete for a given lock $L$.  
Initially, at time $1$, $A$ is the owner, executing in the critical section,
the entry segment is empty and the arrival segment consists of 
$B$ then $C$ then $D$ then $E$ ($B+C+D+E$).  
The non-critical section is empty, so when a thread releases the lock, 
it immediately recirculates and tries to reacquire.  
$A$ completes the critical section and invokes \Release{}, 
which, as the entry segment is empty, reverts to and detaches the arrival segment of $B+C+D+E$ and
moves those threads en-masse to entry segment, and then passes ownership to the head of the entry segment, $B$.  
$A$ recirculates, calls \Acquire{} again, and emplaces itself on the arrival segment, reflecting the state at time $2$.  
Next, $B$ releases the lock, and passes ownership to $C$.  
$B$ then calls \Acquire{} and prepends itself to the arrival segment stack, which now contains
$B+A$, as shown at time 3.  
$C$ releases the lock and conveys ownership to the head of the entry segment, $D$.
$C$ then recirculates and pushes itself onto the arrival segment, which now holds $C+B+A$ at time $4$.  
$D$ releases $L$, cedes ownership to the head of the entry segment, $E$ and then recirculates,
adding itself to the arrival segment, now containing $D+C+B+A$, at time $5$. 
$E$ calls \Release{} and, as the entry segment is empty, $E$ detaches the arrival segment of $D+C+B+A$, shifting
those threads into the arrival segment, and then enables $D$.  $E$ recirculates and 
joins the arrival segment, leaving the configuration as seen at time $6$.  
$D$ releases the lock and passes ownership to $C$ and $D$ then joins the arrival segment, as shown at time $7$. 
$C$ releases $L$ and conveys ownership to $B$, and then prepends itself to the arrival segment, leaving
the state as shown at time $8$. 
$B$ releases $L$, enables $A$, and then joins the arrival segment, leaving the state as shown at time $9$.  
The states at times $1$ and $9$ are identical, so the admission schedule repeats with a period length 
of $8$ steps.  


\subsection{Long-term Admission Unfairness}
While there is no long-term starvation, within the admission cycle $ABCDEDCB$ 
we see that $A$ and $E$ are admitted just once while the others thread are admitted twice, 
which can manifest as long-term relative unfairness between 
the participating threads.  While not a perfect
palindrome, we say such a schedule is \emph{palindromic}
and note that the worst case admission unfairness that might manifest by virtue of the 
lock admission policy is $2X$ -- assuming constant offered load, the most favored thread
can be admitted no more than twice as often as the most ``unlucky'' thread.
In general we find a bimodal distribution of progress over the participating threads. 

\subsection{Cache residency} 
In addition to simple admission, the palindromic schedule introduces an interesting
secondary effect related to residual cache residency.  
We assume a simple system model where all threads circulating over a lock access the same
shared last-level cache (LLC).

While threads are waiting, their residency in the LLC undergoes  decay because of
the actions of the other threads executing in the critical section or their respective non-critical sections.
Arguably, a simple round-robin FIFO FCFS admission schedule results in the worst aggregate miss rate  
and thus the worst overall throughput.  
\footnote{A simple analogy is sequential single-threaded code where an application needs
to iterate over all elements of an array or linked list.
A naive approach is to simply access the elements in ascending order until
reaching the end, and then repeat, yielding a robin-robin order.  
Taking residual cache residency in account, however it is better to alternate 
ascending then descending orders -- akin to a  classic \emph{elevator seek} or \emph{boustrophedonic} 
order -- which yields a palindrome access pattern.  
Such cache-friendly alternating direction optimizations are well known and appear in various 
sorting algorithms \cite{GnomeSort,CockTailSort}}.
In contrast, the palindromic admission order can result in a better aggregate miss rate in the LLC. 
Yuan et al. \cite{Yuan-Locality} use the term ``sawtooth'' for the pure palindrome schedule.  
In our case, the ``elements'' are threads contending for a lock.  
Various threads, however,  can experience persistent disparate LLC miss rates, yielding
a different form of unfairness.  
  


\begin{table} [h]
\tiny
\centering
\begin{tabular}{lllll}
\toprule
Time & Owner & Entry Segment & Arrival Segment &  \\ 


\midrule
1   & A     & -        & B+C+D+E  &        \\
2   & B     & C+D+E    & A        &        \\
3   & C     & D+E      & B+A      &        \\
4   & D     & E        & C+B+A    &        \\ 
5   & E     & -        & D+C+B+A  &        \\
6   & D     & C+B+A    & E        &        \\
7   & C     & B+A      & D+E      &        \\ 
8   & B     & A        & C+D+E    &        \\
9   & A     & -        & B+C+D+E  &        \\

\midrule[\heavyrulewidth]
\bottomrule
\end{tabular}%
\caption{Example of a palindromic admission schedule}\label{Table:Palindromic}
\end{table}



\subsection{Mitigation} 
If desired, we can mitigate the unfairness from the effects above in a number of ways.
A simple and expedient approach is to stochastically disrupt or perturb the repeating cycle, 
which reestablishes statistical long-term fairness.  

A viable technique is 
for incoming owners, having just acquired the lock, to run a thread-local Bernoulli trial,
and based on the outcome, occasionally defer and immediately cede ownership to the next element 
in the entry segment, and propagate a reference to its wait element through the entry segment,
where it will percolate to the tail, and eventually be re-granted ownership.
This modification does not abrogate or otherwise violate our bypass guarantees or imperil anti-starvation
as the reordering is strictly intra-segment.  
This particular approach, while expedient, surrenders the constant-time arrival property as
a thread may need to wait in two phases during one acquisition episode, 
once to acquire ownership, and a 2nd time to be regranted ownership after it abdicated.  

More generally, if we simply pick random elements, without replacement,
from the entry segment for succession, we still retain the desirable population 
bounded anti-starvation property, statistically avoid long-term admission unfairness and 
cache residency fairness, and continue to enjoy aggregate miss rates (and throughput) that
on average are the same or better than classic FIFO.  Crucially, as all such reordering is
intra-segment, we preserve our bounded bypass property.

\section{Conclusion} 

\Invisible{claim, novel, novelty, contribution} 


\recipro{} has a fully constant-time doorway phase 
and a constant-time \Release{}, is practical and fulfills the criteria
for general purpose locking. 
\recipro{} is also \emph{architecturally informed} as it works well with recent developments in modern NUMA 
cache-coherent communication fabrics.

In the future, we plan on exploring the ``coherence traffic reduction'' optimization (CTR), from HemLock, with \recipro.
Relatedly, using the Intel \texttt{CLDEMOTE} instruction or non-temporal stores to covey ownership may also
act to reduce coherence traffic.  
Using modern architecturally enhanced waiting mechanisms such
as \texttt{MONITOR-MWAIT} (Intel) or \texttt{WFE} (ARM) may also prove useful for situations where a thread 
waits for a local flag to toggle and then resets the flag,  as we can instead wait for invalidation of a cache line 
and then use an atomic exchange to try to reset the flag, avoiding intermediate upgrades from MESI/MOESI \emph{shared} to
\emph{modified} states.  
Finally, we are exploring techniques to impose long-term statistical admission fairness by means
of randomizing or perturbing the order of the entry segment in order to break repeating 
palindromic admission schedules.

Hemlock trades off improved space complexity against the cost of higher remote memory 
reference (RMR) complexity. 
Hemlock is exceptionally simple with short paths, and avoids the dependent loads and
indirection required by CLH or MCS to locate queue nodes.  
The contended handover critical path is extremely short --
the unlock operator conveys ownership to the successor in an expedited fashion.  
Despite being compact, it provides 
local spinning in common circumstances and scales better than Ticket Locks.  
Instead of traditional queue elements, as found in CLH and MCS, we use
a per-thread shared singleton element.  
Finally, Hemlock is practical and readily usable in real-world lock implementations. 

\begin{acks}                          
We thank Peter Buhr at the University of Waterloo for access to their ARMv8 system.  
\end{acks}

\bibliography{reciprocating}

\newpage        
\appendix
\newpage 

\section{Naming} 
The name \emph{Reciprocating Locks} arises from the following analogy with a piston-based compressor or
reciprocating pump.  The pump's cylinder has distinct intake and exhaust ports and the piston is initially positioned at the 
bottom of the cylinder.
Threads intending to acquire the lock -- say, $A$, $B$ and $C$ where $A$ was the first to arrive and $C$ last -- 
arrive at the intake port and wait there.
On the intake cycle, the piston pulls those threads from the intake port into the cylinder body. 
On the subsequent exhaust cycle, the piston expels the threads through the exhaust port.  Passing through the 
exhaust port is analogous to passing through the lock's critical section.  Like packets of air or gas, the
threads are expelled from the cylinder through the exhaust port in an order reversed from that
in which they originally arrived : $C$ then $B$ then $A$. 
The \emph{entry segment} corresponds to the current cylinder contents and
the \emph{arrival segment} reflects threads that reside in the intake manifold, before entering the cylinder.  
The non-critical section is where threads circulate back from the exhaust to the intake.
The intake phase is analogous to detaching the current arrival segment and shifting those threads into the entry segment.

\newpage 
\section{On-stack allocation of Wait Elements} 
As described above, our implementation places wait elements in thread-local storage. 
This simplifies reasoning about memory correctness as the tenure and lifespan of thread-local 
storage is the same as that as the associated thread. 
We observe, however, that in many environments, wait elements can also be safely allocated on-stack, 
in the activation frame of \Acquire{}.  
This approach reduces memory usage to $K*E + L*B$ where $K$ is the number of \emph{waiting} threads, 
$E$ is the size of the waiting element in the stack frame, $L$ is the number of currently extant locks, 
and $S$ is the size of the lock body.  

At a conceptual level, the element on which threads spin requires a short lifespan (tenure) and is 
only required to exist and remain in scope for the duration of the \Acquire{} operation, 
and as such can be allocated in the \Acquire{} function's frame.  
This is more convenient in terms of lifecycle than CLH or MCS.  

But, in some circumstances -- zombies -- we also use the address of an element as an end-of-segment marker.
In this case the address escapes the frame as it passed through the \texttt{Gate} field. 
Specifically, the address of the wait element has a longer lifespan than the wait element itself,
and the address persists even after the wait element has fallen out of scope.  (Note that we 
use address-based comparisons to detect the end-of-segment, but never actually reference
the defunct wait elements).  

An address escaping its frame or scope, and having a longer tenure or lifespan that its referent, 
is considered ``undefined behavior'' in C++.  In practice, though, we believe the technique is viable.
If a given virtual address stack address remains associated with a given thread for the lifetime 
of the thread, as is the common case for pthreads environments, then the approach is safe.  
We note, however, that such on-stack allocation might not be safe under exotic non-standard thread 
models where multiple lightweight threads can run or be ``mounted'' on a given stack at different 
points in time, and stack addresses do not map to or otherwise convey thread identity in a stable fashion.
If this particular aspect is a concern, the implementation could simply opt to adhere to our basic design
and place the wait element  into thread-local storage (TLS), which completely obviates the issue.  

We also note that modern compilers can detect such behavior, and generate warnings, and in some cases,
to help protect against undefined behavior, will actively annul references that can be statically
determined to fall out of scope.  

As a thread can wait on at most on
lock at a time, and a wait element is only needed for the duration of an \Acquire{} operation,
using a singleton in thread-local storage is entirely sufficient.
We are using $E$'s address for addressed-based comparisons as a distinguished marker to indicate 
the logical end-of-segment.  $E$ itself will not be subsequently accessed.

\Invisible{UB; Nasal Demons; halt-and-catch-fire}


\newpage 
\section{Potential Throughput Benefits from Palindromic Schedule Order} 

We assume a simple system model where all threads circulating over a lock access the same 
shared last-level cache (LLC). 
While threads are waiting, their residency in the LLC undergoes decay -- usually modelled as exponential -- because of
the actions of the other threads executing in the critical section or their respective non-critical sections. 
We consider a true repeating palindrome admission schedule, $ABCDE-EDCBA$, as compared to
the FIFO schedule of $ABCDE-ABCDE$.  
Applying a simplistic decay model, when a thread ceases waiting and takes 
ownership of the lock, it incurs a ``cache reload transient''\cite{sigmetrics86-stone} where it
suffers a burst of cache misses as it reprovisions the LLC with its own previously displaced
private data.  The residual residency fraction can be approximated as $Residual(T) = exp(-T*\lambda)$ where
$T$ is the sojourn or waiting time -- the number of quanta since the thread last ran -- and $\lambda$ 
parameterizes the decay rate.  
$\lambda$ is usually expressed as \emph{half-life}. 
As $Residual$ is a convex function, we can employ Jensen's inequality \cite{boyd2004convex} as follows. 
Taking thread $B$ as a specific example, its waiting times under the FIFO schedule is always 4 time units and
under the palindrome schedule the waiting time alternate 2-6-2-6 etc.  
The average waiting time is the same under both schedules,
but the average residual LLC residency when the thread resumes is the same or better under 
the palindrome schedule as $Residual(2)+Residual(6) \ge Residual(4)+Residual(4)$.  
In fact, each and every thread will have the same or better residual fraction under 
the palindrome schedule than under FIFO
although there will be disparity of benefits over the set of threads within the palindrome schedule.  

Intuitively, as the decay process is exponential in nature, the retained residency benefits accrued 
by the relatively short gap outweigh the decay penalty of the subsequent longer gap found in 
the palindrome schedule.

The overall aggregate miss rate for the palindrome schedule, 
as computed over all the threads, will be less than that in the round-robin FIFO schedule, yielding
better overall throughput.  
(Higher residency fractions implies reduced miss rates and better performance). 
Specifically, The palindrome schedule enjoys better overall aggregate LLC miss rates and 
throughput than a simple repeating round-robin FIFO schedule of $ABCDE-ABCDE$.  
And in fact the FIFO schedule is pessimal for aggregate miss rate if we require equal fairness 
as measured over two back-to-back cycles.  


The same palindrome schedule effects on cache residency also apply by analogy to page working sets.
Also by analogy, a simple uniprocessor scheduler would admit better cache residency
using a palindrome schedule as compared to a classic FIFO round-robin schedule.


We call out an analogy in simple sequential single-threaded code where an application needs
to iterate over all elements of an array or linked list.
A naive approach is to simply access the elements in ascending order until
reaching the end, and then repeat, yielding a robin-robin order.  
Taking residual cache residency in account, however it is better to alternate 
ascending then descending orders -- akin to a  classic \emph{elevator seek} or \emph{boustrophedonic} 
order -- which yields a palindrome access pattern.  
Such cache-friendly alternating direction optimizations are well known and appear in various 
sorting algorithms \cite{GnomeSort,CockTailSort}.
Yuan et al. \cite{Yuan-Locality} use the term ``sawtooth'' for the palindrome schedule.  
In our case, the ``elements'' are threads contending for a lock.  


Considering a palindrome lock admission order of $ABCDE-EDCBA-ABCDE...$,  for instance, 
we will have fair admission over the long term, but threads $A$ and $E$ will enjoy persistently lower shared cache miss
rates than the other threads, imposing a different form of unfairness related to residual cache residency.
Crucially, under the palidrome schedule, threads can incur disparate cache hits rates, 
reflecting a form of long-term cache-based unfairness, even if the the admission is long-term fair.  
The overall aggregate miss rate for the palindrome schedule, however, as
computed over all the threads, will be less than that in the round-robin FIFO schedule.  
We note in passing that a random admission order is statistically long-term fair for both admission
frequency and cache residency, and also will display a lower aggregate cache miss rate than FIFO.  
The same effects that apply to cache residency also have analogs in page working sets.  

\Invisible{As a corallary to the above, if we assume a set of threads equitably but otherwise 
distributed in a random fashion over a set of NUMA nodes, a palindrome admission order exhibits, 
on average, fewer NUMA \emph{lock migrations} than does a round-robin (FIFO) order,
making palindrome admission innately NUMA-friendly without being specifically NUMA-aware. } 

\Invisible{apply by analogy to} 
\Invisible{Palindrome-adjacent}

\newpage 
\section{Locking APIs}

Modern C++ locking constructs such as \texttt{std::scoped\_lock} and \texttt{std::lock\_guard},
by means of the \emph{Resource Acquisition is Initialization(RAII)}\cite{RAII} idiom, where
the constructor acquires the lock and the destructor releases the lock, 
manage to avoid explicit \texttt{lock} and \texttt{unlock} calls in application code.
This same design pattern readily supports underlying lock primitives that require context to be passed.  
Specifically, we can pass extra context through additional fields in the RAII wrapper classes. 
Likewise, locking interfaces that specify the critical section as a C++ lambda also allow such latitude.  
Interfaces that use scoped locking, such as Java's ``synchronized" construct, permit
the implementation to trivially pass information from the underlying \texttt{lock} cite to the
corresponding \texttt{unlock}.  


If an implementation is forced to use a legay interface, such as that exposed via \texttt{pthread\_\allowbreak{}mutex\_\allowbreak{}lock}
and \texttt{pthread\_\allowbreak{}mutex\_\allowbreak{}unlock}, or C++ \texttt{std::BasicLockable}, but the lock algorithm
is not \emph{context-free}, then we can resort to passing context information through extra fields
in the lock body, which are protected by the lock itself, and written and read only by the lock holder. 
This increases the size of the lock body and accesses to those fields can induce extra coherence traffic. 
Another approach is to keep track of held locks in thread-local storage (TLS) and convey the information in that fashion.  
Most locking algorithms that are not innately \emph{context-free} can be transformed to become \emph{context-free}
through such techniques. 
We note that MCS and CLH also also must pass additional contextual information from \Acquire{} to \Release{}, 
so this concern is not specific to \recipro{}.  

More modern interfaces,as noted above, may confer various advantages and obviate some
of the concerns associated with passing context. 
We hope these observations serve to inform API designers, in that more flexible and liberal locking
APIs may more easily accomodate a wider array of underlying lock algorithms, and afford more latitude to 
implementors.  


\Invisible{Lesson, Informs, Instructs, Guides}


\newpage 
\section{Simplified Example}
In Listing-\ref{Listing:WFTC3} we show, for purposes of explication,
a slightly simplified form of \recipro{}, which is both easy to understand and implement.  
We recommend that implementors start with this version.  
This variant passes the identity of the successor (\texttt{succ}) as context.
The identity of the \emph{terminus} end-of-segment address is encoding via
a global variable, \texttt{eos}, in the lock body.  To reduce coherence traffic
and false sharing between \texttt{eos} and \texttt{Arrivals}, we sequester and isolate
\texttt{eos} as the sole occupant of its cache sector.  
Crucially, \texttt{eos} is not written during steady-state sustained contention, so
does not tend to generate coherence misses in that circumstance.  

\Invisible{Exemplar; explainer} 

\onecolumn
\lstset{caption={Reciprocating Locks -- Simplified Starting Point}}
\lstset{label={Listing:WFTC3}}
\begin{lstlisting}[mathescape=true,escapechar=\%]
struct ReciprocatingLock { 
 %\IRule% struct WaitElement { 
 %\IRule%  %\IRule% std::atomic<int> Gate {0} ; 
 %\IRule% } ; 
 %\IRule%  
 %\IRule% // CONSIDER: packed or dense layout for Arrivals vs eos
 %\IRule% // Arrival word : 
 %\IRule% std::atomic <WaitElement *> Arrivals {nullptr} ; 
 %\IRule% // Terminus end-of-segment sentinel : 
 %\IRule% std::atomic <WaitElement *> eos alignas(128)  {nullptr} ;           
 %\IRule%  
 %\IRule% static inline const auto NEMO = (WaitElement *) uintptr_t(1) ; 
 %\IRule% static inline constinit thread_local WaitElement E alignas(128) {0} ; 
 %\IRule%  
 public : 
 %\IRule% inline auto %\colorbox{yellow!50}{operator+}%(std::invocable auto && csfn) %$\rightarrow$% void {
 %\IRule%  %\IRule% // %\underline{Acquire Phase}% ...
 %\IRule%  %\IRule% E.Gate.store (0,  std::memory_order_release) ;
 %\IRule%  %\IRule% auto succ = Arrivals.exchange (&E) ; 
 %\IRule%  %\IRule% assert (succ %$\neq$% &E) ; 
 %\IRule%  %\IRule% if (succ == nullptr) { 
 %\IRule%  %\IRule%  %\IRule% // fast-path uncontended aquire 
 %\IRule%  %\IRule%  %\IRule% eos.store (&E, std::memory_order_release) ; 
 %\IRule%  %\IRule% } else { 
 %\IRule%  %\IRule%  %\IRule% // Coerce NEMO to nullptr 
 %\IRule%  %\IRule%  %\IRule% // succ will be our successor when we subsequently release
 %\IRule%  %\IRule%  %\IRule% succ = (WaitElement *) (uintptr_t(succ) & ~1) ;
 %\IRule%  %\IRule%  %\IRule% assert (succ %$\neq$% &E) ;
 %\IRule%  %\IRule%  %\IRule%  
 %\IRule%  %\IRule%  %\IRule% // slow path : contention -- waiting phase
 %\IRule%  %\IRule%  %\IRule% while (E.Gate.load (std::memory_order_acquire) == 0) { 
 %\IRule%  %\IRule%  %\IRule%  %\IRule% Pause() ; 
 %\IRule%  %\IRule%  %\IRule% }
 %\IRule%  %\IRule%  %\IRule%  
 %\IRule%  %\IRule%  %\IRule% // Check for EOS-termimated entry segment chain 
 %\IRule%  %\IRule%  %\IRule% // On match : annul succ and reset global eos variable
 %\IRule%  %\IRule%  %\IRule% // CRITICAL : Note that eos does NOT change under sustained 
 %\IRule%  %\IRule%  %\IRule% // contention after we reach steady state !  
 %\IRule%  %\IRule%  %\IRule% // The global "eos" field becomes stable 
 %\IRule%  %\IRule%  %\IRule% auto veos = eos.load (std::memory_order_acquire) ; 
 %\IRule%  %\IRule%  %\IRule% assert (veos %$\neq$% &E) ; 
 %\IRule%  %\IRule%  %\IRule% assert (veos %$\neq$% nullptr) ; 
 %\IRule%  %\IRule%  %\IRule% if (succ == veos) { 
 %\IRule%  %\IRule%  %\IRule%  %\IRule% succ = nullptr ; 
 %\IRule%  %\IRule%  %\IRule%  %\IRule% eos.store (NEMO, std::memory_order_release) ; 
 %\IRule%  %\IRule%  %\IRule% }
 %\IRule%  %\IRule% }
 %\IRule%  %\IRule%  
 %\IRule%  %\IRule% // Critical section expressed as lambda 
 %\IRule%  %\IRule% // We pass "succ" as context from Acquire to corresponding Release
 %\IRule%  %\IRule% // The Release phase does NOT access eos 
 %\IRule%  %\IRule% // Access to global eos is constrained to Acquire
 %\IRule%  %\IRule% // &E is implicit context 
 %\IRule%  %\IRule% EnterCS: 
 %\IRule%  %\IRule% assert (Arrivals.load() %$\neq$% nullptr) ; 
 %\IRule%  %\IRule% csfn() ;
 %\IRule%  %\IRule% assert (Arrivals.load() %$\neq$% nullptr) ; 
 %\IRule%  %\IRule%  
 %\IRule%  %\IRule% // %\underline{Acquire Phase}% ...
 %\IRule%  %\IRule% // Case : entry list populated -- appoint successor from entry list
 %\IRule%  %\IRule% if (succ %$\neq$% nullptr) {
 %\IRule%  %\IRule%  %\IRule% assert (succ%$\rightarrow$%Gate.load(std::memory_order_acquire) == 0) ; 
 %\IRule%  %\IRule%  %\IRule% succ%$\rightarrow$%Gate.store (1, std::memory_order_release) ;
 %\IRule%  %\IRule%  %\IRule% return ;
 %\IRule%  %\IRule% }
 %\IRule%  %\IRule%  
 %\IRule%  %\IRule% // Case : entry list and arrivals are both empty
 %\IRule%  %\IRule% // try fast-path uncontended unlock 
 %\IRule%  %\IRule% auto k = Arrivals.load (std::memory_order_acquire) ;
 %\IRule%  %\IRule% if (k == &E || k == NEMO) { 
 %\IRule%  %\IRule%  %\IRule% if (Arrivals.compare_exchange_strong (k, nullptr)) return ; 
 %\IRule%  %\IRule% }
 %\IRule%  %\IRule%  
 %\IRule%  %\IRule% // Case : entry list is empty and arrivals is populated 
 %\IRule%  %\IRule% // New threads have arrived and pushed themselves onto the 
 %\IRule%  %\IRule% // arrival stack. 
 %\IRule%  %\IRule% // We now detach that segment, shifting those arrivals to 
 %\IRule%  %\IRule% // become the next entry segment 
 %\IRule%  %\IRule% Arrivals.exchange(NEMO)%$\rightarrow$%Gate.store (1, std::memory_order_release) ; 
 %\IRule% } 
} ; 

\end{lstlisting}
\twocolumn

\newpage 
\section{Variations}

All the implementations in this section employ a \emph{double swap} arrival for
uncontended operations.  Arriving threads, as usual, use an atomic exchange (swap)
to first install the address of their waiting element, $E$.  If the exchange
returned \emph{nullptr}, then the thread has gained ownership.  In that case,
to try to avoid $E$ becoming submerged, they immediately exchange \texttt{LOCKEDEMPTY}
into the \texttt{Arrival} word.  If that second exchanged returned $E$, then the 
\emph{double swap} was successful, and the thread can immediately enter the critical section.
If other threads arrived between between two exchange operations, however, then 
$E$ become submerged and the second swap detached a new arrival segment, with $E$
residing at the ``far'' end.  Our thread needs to recover from that condition.  

In Listing-\ref{Listing:WFvCB}, in the event of an arrival race, our thread,
which is still currently the owner, simply cedes ownership to the most recently arrived 
thread on the newly detached segment, identified via $R$ (Line-56) and then proceeds to wait.  
This approach is extremely simple and avoids any need to use or convey end-of-segment information.  
The only context that needs to be passed to the \Release{} phase is the address of the \emph{successor}.  
This approach also preserves the desirable thread-specific bounded bypass property, but, arguably,
it has poorer progress properties and is no longer constant-time as ownership needs to 
be relayed for the victim to $R$.  

In practice we find the arrival race between the two swaps to be rare, likely as the window
of vulnerability is short, and because it takes time for the coherent interconnect 
to re-arbitrate the cache line between processors.
Also, the race can only occur at the \emph{onset} of contention, when the first arriving thread found
the lock not held and then other threads arrived in quick succession.

As a trivial and racy counter-measure, to reduce the rate of \emph{double swap} arrivals, 
in \Acquire{}, we can first load the arrival word, and, if it is \texttt{nullptr}, conditionally 
attempt an atomic CAS to opportunistically try to swing the word to \texttt{LOCKEDEMPTY}.  
If the CAS is successful, the thread can return immediately, otherwise we let control fall through
into the swap path.  This may act to reduce the incidence of double-swap operations for mostly
uncontended locks.  

In theory the \emph{double swap} serves to increase theoretical RMR complexity
complexity \cite{stoc08-attiya,podc02-anderson}.   
In practice, however, as the lock is uncontended, the underlying cache line tends to
remain in local \emph{modified} state in arriving thread's cache, 
so the second exchange incurs very little additional cost and no additional coherence misses. 
As noted above, the double swap is executed only absent contention, where the arriving thread 
found the lock in \emph{unlocked} state.  
Under sustained contention, we avoid the \emph{double swap}.  

\Invisible{The variation show in Listing-\ref{Listing:WFvCB} avoids the use
of the end-of-segment, yielding a simplified algorithm.  
This also reduces context that needs to be passed from \Acquire{} to \Release{} to
just the identity, if any, of the successor. 
If additional threads raced and arrived and pushed onto the arrival segment in 
the window between the two exchange operations, the current owner simply abdicates and relays 
ownership to the first thread in the newly detached entry list, and then proceeds to wait.  
While simpler, we believe this variant has poorer progress properties, as, when
the race manifests, ownership needs to relay through the race victim thread.} 

In this particular variation we can safely allocate the waiting elements on stack, if desired,
as the address of the element $E$ never escapes the \Acquire{} phase, even if 
\Acquire{} and \Release{} are implemented as different functions.  


In Listing-\ref{Listing:RotaryWFDU} we show another variation that uses only one atomic 
\texttt{fetch-and-add(1)} in the \Release{} phase.  This form significantly
alters the encoding of the \texttt{Arrival} word, using the low-order two bits as a tag. 
This form avoids the need for an explicit \texttt{LOCKEDEMPTY} distinguished value,
and also avoids the need to convey end-of-segment addresses through the chain. 
The only context that needs to be passed to the \Release{} phase is the address
of the successor on the detached entry segment.


Listing-\ref{Listing:RotaryWFF} shows a variation where a thread retains ownership even if other threads 
arrived in the \emph{double swap} or \texttt{exchange}-\texttt{fetch-and-add} window.  In this case the second atomic
in the putative uncontended arrival path -- be it \texttt{exchange} or \texttt{fetch-and-add} -- detached
an arrival segment identified via $R$ (Line-40).  The owner records $R$ as its successor, and also passes
the address of $E$ through $R$ towards the tail of entry segment (Line-59).  In this particular implementation, the
end-of-segment \texttt{eos} field is used only at the onset of contention and only when the race occured in the
arrival window, and \texttt{eos} is always otherwise \emph{nullptr}.  In addition, the use of \texttt{eos} is 
encapsulated and restricted to the \Acquire{} phase, and \texttt{eos} does not need to communicated as context
to the corresponding \Release{} phase, resulting in a more simple and easy to understand algorithm.   

We note that approaches in Listing-\ref{Listing:WFvCB} and Listing-\ref{Listing:RotaryWFF} can easily be combined
into a form that avoids the use of atomic \texttt{fetch-and-add}, yielding Listing-\ref{Listing:WFvBD}.   


\newpage 
\onecolumn

\lstset{caption={Reciprocating Locks -- Relay}}
\lstset{label={Listing:WFvCB}}
\begin{lstlisting}[mathescape=true,escapechar=\%]
struct ReciprocatingLock { 
 %\IRule% struct WaitElement { 
 %\IRule%  %\IRule% std::atomic<int> Gate alignas(128) {0} ; 
 %\IRule% } ; 
 %\IRule% 
 %\IRule% Atomic <WaitElement *> Arrivals {nullptr} ; 
 %\IRule% static inline const auto LOCKEDEMPTY = 
 %\IRule%  %\IRule% (WaitElement *) uintptr_t(1) ; 
 %\IRule% 
 public : 
 %\IRule% inline auto %\colorbox{yellow!50}{operator+}%(std::invocable auto && csfn) %$\rightarrow$% void {
 %\IRule%  %\IRule% // %\underline{Acquire Phase}% ...
 %\IRule%  %\IRule% // Can place E either on-stack or in TLS
 %\IRule%  %\IRule% // Consider alignas(128) 
 %\IRule%  %\IRule% WaitElement E {} ; 
 %\IRule%  %\IRule% auto succ = (WaitElement *) nullptr ; 
 %\IRule%  %\IRule% auto tail = Arrivals.exchange (&E) ; 
 %\IRule%  %\IRule% assert (tail %$\neq$% &E) ; 
 %\IRule%  %\IRule% // fast-path uncontended acquire 
 %\IRule%  %\IRule% if (tail == nullptr) { 
 %\IRule%  %\IRule%  %\IRule% auto R = Arrivals.exchange (LOCKEDEMPTY) ; 
 %\IRule%  %\IRule%  %\IRule% assert (R %$\neq$% nullptr && R %$\neq$% LOCKEDEMPTY) ; 
 %\IRule%  %\IRule%  %\IRule% if (R == &E) goto EnterCS ; 
 %\IRule%  %\IRule%  %\IRule% // Cede -- Abdicate and fall thru into waiting phase ...
 %\IRule%  %\IRule%  %\IRule% // Other threads arrived and pushed onto L%$\rightarrow$%Arrivals in
 %\IRule%  %\IRule%  %\IRule% // the Exchange-Exchange window, above
 %\IRule%  %\IRule%  %\IRule% // Our &E is now burried in the arrival stack.
 %\IRule%  %\IRule%  %\IRule% // We expect this scenario to be relatively rare. 
 %\IRule%  %\IRule%  %\IRule% //
 %\IRule%  %\IRule%  %\IRule% // We recover and compensate by passing or diverting 
 %\IRule%  %\IRule%  %\IRule% // ownership directly to the thread associated with R.  
 %\IRule%  %\IRule%  %\IRule% // Our own element, E, which resides within the now-detached 
 %\IRule%  %\IRule%  %\IRule% // stack, will eventually gain ownership via natural 
 %\IRule%  %\IRule%  %\IRule% // succession. 
 %\IRule%  %\IRule%  %\IRule% // R is the most-recently-arrivated thread (at the time 
 %\IRule%  %\IRule%  %\IRule% // of the 2nd exchange, above) and is the head of that 
 %\IRule%  %\IRule%  %\IRule% // detached stack. 
 %\IRule%  %\IRule%  %\IRule% //
 %\IRule%  %\IRule%  %\IRule% // This form arguably does _not provide a classic bounded 
 %\IRule%  %\IRule%  %\IRule% // wait-free transfer of ownership, as, in this path, 
 %\IRule%  %\IRule%  %\IRule% // ownership needs to pass transiently through this thread 
 %\IRule%  %\IRule%  %\IRule% // as an intermediary before reaching R.  
 %\IRule%  %\IRule%  %\IRule% // We relay ownership immediately to R. 
 %\IRule%  %\IRule%  %\IRule% // Succession and progress depend on this thread making 
 %\IRule%  %\IRule%  %\IRule% // progress to enable the actual successor, R.  
 %\IRule%  %\IRule%  %\IRule% //
 %\IRule%  %\IRule%  %\IRule% // And in general, if a lock acquisition episode has 2 
 %\IRule%  %\IRule%  %\IRule% // distinct waiting phases or modes for a given thread, 
 %\IRule%  %\IRule%  %\IRule% // then we don't have have a constant-time doorway.  
 %\IRule%  %\IRule%  %\IRule% // 
 %\IRule%  %\IRule%  %\IRule% // And while simpler, this approach can generate
 %\IRule%  %\IRule%  %\IRule% // more coherent communication traffic as ownership bounces 
 %\IRule%  %\IRule%  %\IRule% // from this thread and then immediately to the intended 
 %\IRule%  %\IRule%  %\IRule% // successor, R.  
 %\IRule%  %\IRule%  %\IRule% assert (R%$\rightarrow$%Gate.load() == 0) ; 
 %\IRule%  %\IRule%  %\IRule% R%$\rightarrow$%Gate.store (1, std::memory_order_release) ; 
 %\IRule%  %\IRule%  %\IRule% // fall through into normal waiting ...
 %\IRule%  %\IRule% }
 %\IRule%  %\IRule%  
 %\IRule%  %\IRule% // Coerce LOCKEDEMPTY to nullptr 
 %\IRule%  %\IRule% // succ will be our successor when we subsequently release
 %\IRule%  %\IRule% succ = (WaitElement *) (uintptr_t(tail) & ~1) ;
 %\IRule%  %\IRule% assert (succ %$\neq$% &E) ;
 %\IRule%  %\IRule% 
 %\IRule%  %\IRule% // slow path : contention -- waiting phase
 %\IRule%  %\IRule% while (E.Gate.load(std::memory_order_acquire) == 0) { 
 %\IRule%  %\IRule%  %\IRule% Pause() ; 
 %\IRule%  %\IRule% }
 %\IRule%  %\IRule%  
 %\IRule%  %\IRule% // Critical section expressed as lambda 
 %\IRule%  %\IRule% // We pass "succ" as context from Acquire to corresponding Release
 %\IRule%  %\IRule% EnterCS: 
 %\IRule%  %\IRule% assert (Arrivals.load() %$\neq$% nullptr) ; 
 %\IRule%  %\IRule% csfn() ;
 %\IRule%  %\IRule% assert (Arrivals.load() %$\neq$% nullptr) ; 
 %\IRule%  %\IRule%  
 %\IRule%  %\IRule% // %\underline{Release Phase}% ...
 %\IRule%  %\IRule% // Case : entry list populated -- appoint successor from entry list
 %\IRule%  %\IRule% // If successor exists on implicit Entrylist, 
 %\IRule%  %\IRule% // just pass ownership to that waiting thread. 
 %\IRule%  %\IRule% if (succ %$\neq$% nullptr) {
 %\IRule%  %\IRule%  %\IRule% assert (succ%$\rightarrow$%Gate.load() == 0) ;
 %\IRule%  %\IRule%  %\IRule% succ%$\rightarrow$%Gate.store (1, std::memory_order_release) ;
 %\IRule%  %\IRule%  %\IRule% return ;
 %\IRule%  %\IRule% }
 %\IRule%  %\IRule%  
 %\IRule%  %\IRule% // Case : entry list and arrivals are both empty
 %\IRule%  %\IRule% // try fast-path uncontended unlock 
 %\IRule%  %\IRule% // Can use either raw CAS or polite LD-CAS 
 %\IRule%  %\IRule% auto k = LOCKEDEMPTY ; 
 %\IRule%  %\IRule% if (Arrivals.compare_exchange_strong (k, nullptr)) return ; 
 %\IRule%  %\IRule%  
 %\IRule%  %\IRule% // Case : entry list is empty and arrivals is populated 
 %\IRule%  %\IRule% // New threads have arrived and pushed themselves onto the 
 %\IRule%  %\IRule% // arrival stack. 
 %\IRule%  %\IRule% // We now detach that segment, shifting those arrivals to 
 %\IRule%  %\IRule% // become the next entry segment 
 %\IRule%  %\IRule% // Decapitate the arrival segment, leaving the lock held.  
 %\IRule%  %\IRule% // Reprovision the entrylist with those arrivals.  
 %\IRule%  %\IRule% auto w = Arrivals.exchange (LOCKEDEMPTY) ;
 %\IRule%  %\IRule% assert (w %$\neq$% nullptr && w %$\neq$% LOCKEDEMPTY) ; 
 %\IRule%  %\IRule% assert (w%$\rightarrow$%Gate.load() == 0) ; 
 %\IRule%  %\IRule% w%$\rightarrow$%Gate.store (1, std::memory_order_release) ;
 %\IRule% } 
} ; 

\end{lstlisting}

\newpage 
\lstset{caption={Reciprocating Locks -- fetch-add}}
\lstset{label={Listing:RotaryWFDU}}
\begin{lstlisting}[mathescape=true,escapechar=\%]
// @  Avoids LOCKEDEMPTY encoding
// @  Arrivals encoding : tagged pointer 
//    E:00 :  Locked   --  Arrival stack populated 
//    E:01 :  Locked   --  Arrival segment logically detached and empty 
//    *:10 :  Unlocked --  Equivalent to nullptr : end-of-segment
//    *:11 :  illegal and undefined 
//    0:00 :  illegal and undefined 
// @  Intentionally designed encoding above to work with atomic fetch_add(1)  
//    fetch_add(1) low-bit values drives state machine 
//    Arrived %$\rightarrow$% detached %$\rightarrow$% unlocked 
// @  Must initialize Arrivals to 0:10 
// @  Very tight paths with reduced branch complexity
//    Retains constant-time wait-free arrival and unlock paths
//    Retains usual Reciprocating admission order
//    Retains thread-specific bounded bypass 
//    Release phase has only one atomic !
 
struct ReciprocatingLock { 
 %\IRule% struct WaitElement { 
 %\IRule%  %\IRule% std::atomic<int> Gate alignas(128) {0} ; 
 %\IRule% } ; 
 %\IRule%  
 %\IRule% std::atomic <WaitElement *> Arrivals { (WaitElement *) uintptr_t(2) } ; 
 %\IRule% static inline constinit thread_local WaitElement E alignas(128) {0} ; 
 %\IRule%  
 %\IRule% static WaitElement * AnnulMarked (WaitElement * v) {
 %\IRule%  %\IRule% auto u = uintptr_t(v) ; 
 %\IRule%  %\IRule% return (WaitElement *) (u & ((u & 1) - 1)) ; 
 %\IRule% }
 %\IRule% 
 %\IRule% // Mark and return MRA element but leave it in emplaced on chain 
 %\IRule% WaitElement * FetchAndMark () { 
 %\IRule%  %\IRule% return (WaitElement *) 
 %\IRule%  %\IRule%  (((std::atomic<uintptr_t> *) &this%$\rightarrow$%Arrivals)%$\rightarrow$%fetch_add(1)) ;      
 %\IRule% } 
 %\IRule%  
 public : 
 %\IRule% inline auto %\colorbox{yellow!50}{operator+}%(std::invocable auto && csfn) %$\rightarrow$% void {
 %\IRule%  %\IRule% // %\underline{Acquire Phase}% ...
 %\IRule%  %\IRule% E.Gate.store (0, std::memory_order_release) ; 
 %\IRule%  %\IRule% auto succ = Arrivals.exchange (&E) ; 
 %\IRule%  %\IRule% assert (succ %$\neq$% &E) ; 
 %\IRule%  %\IRule% assert ((uintptr_t(succ) & 3) %$\neq$% 3) ; 
 %\IRule%  %\IRule% assert (uintptr_t(succ) %$\neq$% 0) ; 
 %\IRule%  %\IRule% if (uintptr_t(succ) & 2) { 
 %\IRule%  %\IRule%  %\IRule% // We are the owner -- uncontended acquisition
 %\IRule%  %\IRule%  %\IRule% // exchange() transitioned Arrivals low-order bits from 
 %\IRule%  %\IRule%  %\IRule% // unlocked to lock state 
 %\IRule%  %\IRule%  %\IRule% auto R = FetchAndMark () ; 
 %\IRule%  %\IRule%  %\IRule% assert ((uintptr_t(R) & 3) == 0) ; 
 %\IRule%  %\IRule%  %\IRule% succ = nullptr ; 
 %\IRule%  %\IRule%  %\IRule% // fast path uncontended arrival ...
 %\IRule%  %\IRule%  %\IRule% if (R == &E) goto EnterCS ; 
 %\IRule%  %\IRule%  %\IRule%  
 %\IRule%  %\IRule%  %\IRule% // New threads arrived within the narrow 
 %\IRule%  %\IRule%  %\IRule% // exchange-fetch_add() window. 
 %\IRule%  %\IRule%  %\IRule% // This should be relatively rare
 %\IRule%  %\IRule%  %\IRule% // Occurs only at the onset of contention
 %\IRule%  %\IRule%  %\IRule% // Just delegate ownership of lock to the 
 %\IRule%  %\IRule%  %\IRule% // most-recently-arrived-thread "R"
 %\IRule%  %\IRule%  %\IRule% // R is now the head of a detached segment
 %\IRule%  %\IRule%  %\IRule% // Our E resides buried at the distal end of that 
 %\IRule%  %\IRule%  %\IRule% // detached segment.  
 %\IRule%  %\IRule%  %\IRule% // succ is effectively null
 %\IRule%  %\IRule%  %\IRule% assert (R %$\neq$% nullptr) ; 
 %\IRule%  %\IRule%  %\IRule% assert (R%$\rightarrow$%Gate.load() == 0) ; 
 %\IRule%  %\IRule%  %\IRule% R%$\rightarrow$%Gate.store (1, std::memory_order_release) ; 
 %\IRule%  %\IRule%  %\IRule% // fall through into waiting 
 %\IRule%  %\IRule% }
 %\IRule%  %\IRule%  
 %\IRule%  %\IRule% succ = AnnulMarked (succ) ; 
 %\IRule%  %\IRule% assert ((uintptr_t(succ) & 3) == 0) ; 
 %\IRule%  %\IRule% assert (succ %$\neq$% &E) ; 
 %\IRule%  %\IRule% while (E.Gate.load(std::memory_order_acquire) == 0) { 
 %\IRule%  %\IRule%  %\IRule% Pause() ; 
 %\IRule%  %\IRule% }
 %\IRule%  %\IRule%  
 %\IRule%  %\IRule% // Critical section expressed as lambda 
 %\IRule%  %\IRule% // pass "succ" as context Acquire to corresponding Release
 %\IRule%  %\IRule% EnterCS: 
 %\IRule%  %\IRule% assert (succ %$\neq$% &E) ; 
 %\IRule%  %\IRule% assert ((uintptr_t(Arrivals.load()) & 2) == 0) ; 
 %\IRule%  %\IRule% assert (uintptr_t(Arrivals.load()) %$\neq$% 0) ; 
 %\IRule%  %\IRule% csfn() ;
 %\IRule%  %\IRule% assert ((uintptr_t(Arrivals.load()) & 2) == 0) ; 
 %\IRule%  %\IRule% assert (uintptr_t(Arrivals.load()) %$\neq$% 0) ; 
 %\IRule%  %\IRule%  
 %\IRule%  %\IRule% // %\underline{Release Phase}% ...
 %\IRule%  %\IRule% if (succ == nullptr) { 
 %\IRule%  %\IRule%  %\IRule% // Mark element but leave it in emplaced on chain 
 %\IRule%  %\IRule%  %\IRule% succ = FetchAndMark () ; 
 %\IRule%  %\IRule%  %\IRule% assert ((uintptr_t(succ) & 2) == 0) ; 
 %\IRule%  %\IRule%  %\IRule% assert (uintptr_t(succ) %$\neq$% 0) ; 
 %\IRule%  %\IRule%  %\IRule% if (uintptr_t(succ) & 1) return ; 
 %\IRule%  %\IRule%  %\IRule%  
 %\IRule%  %\IRule%  %\IRule% // We just detached fresh arrivals 
 %\IRule%  %\IRule%  %\IRule% assert ((uintptr_t(succ) & 3) == 0) ; 
 %\IRule%  %\IRule%  %\IRule% assert (succ %$\neq$% nullptr) ; 
 %\IRule%  %\IRule% }
 %\IRule%  %\IRule%  
 %\IRule%  %\IRule% assert (succ%$\rightarrow$%Gate.load() == 0) ;
 %\IRule%  %\IRule% succ%$\rightarrow$%Gate.store (1, std::memory_order_release) ; 
 %\IRule% } 
} ;


\end{lstlisting}


\newpage
\lstset{caption={Reciprocating Locks -- Simplified}}
\lstset{label={Listing:RotaryWFF}}
\begin{lstlisting}[mathescape=true,escapechar=\%]
struct ReciprocatingLock { 
 %\IRule% struct WaitElement { 
 %\IRule%  %\IRule% std::atomic <int> Gate alignas(128) {0} ; 
 %\IRule%  %\IRule% std::atomic <WaitElement *> eos {nullptr} ; 
 %\IRule% } ; 
 %\IRule%  
 %\IRule% std::atomic <WaitElement *> Arrivals { (WaitElement *) uintptr_t(2) } ; 
 %\IRule%  
 %\IRule% // Variations for E placement : 
 %\IRule% // @  auto on-stack in operator()+
 %\IRule% // @  class scoped : static inline thread_local
 %\IRule% // @  method local : static thread_local
 %\IRule% static inline constinit thread_local WaitElement E alignas(128) {0, nullptr} ; 
 %\IRule%  
 %\IRule% static WaitElement * AnnulMarked (WaitElement * v) {
 %\IRule%  %\IRule% auto u = uintptr_t(v) ; 
 %\IRule%  %\IRule% return (WaitElement *) (u & ((u & 1) - 1)) ; 
 %\IRule% }
 %\IRule%  
 %\IRule% // Mark and return MRA element but leave it in emplaced on chain 
 %\IRule% WaitElement * FetchAndMark () { 
 %\IRule%  %\IRule% return (WaitElement *) 
 %\IRule%  %\IRule%  %\IRule% (((std::atomic<uintptr_t> *) &this%$\rightarrow$%Arrivals)%$\rightarrow$%fetch_add(1)) ;      
 %\IRule% } 
 %\IRule%  
 public : 
 %\IRule% inline auto %\colorbox{yellow!50}{operator+}%(std::invocable auto && csfn) %$\rightarrow$% void {
 %\IRule%  %\IRule% // %\underline{Acquire Phase}%
 %\IRule%  %\IRule% E.eos.store  (nullptr, std::memory_order_release) ; 
 %\IRule%  %\IRule% E.Gate.store (0      , std::memory_order_release) ; 
 %\IRule%  %\IRule% auto succ = Arrivals.exchange (&E) ; 
 %\IRule%  %\IRule% assert (succ %$\neq$% &E) ; 
 %\IRule%  %\IRule% assert ((uintptr_t(succ) & 3) %$\neq$% 3) ; 
 %\IRule%  %\IRule% assert (uintptr_t(succ) %$\neq$% 0) ; 
 %\IRule%  %\IRule% if (uintptr_t(succ) & 2) { 
 %\IRule%  %\IRule%  %\IRule% // We are the owner -- uncontended acquisition
 %\IRule%  %\IRule%  %\IRule% // exchange() transitioned Arrivals low-order bits from 
 %\IRule%  %\IRule%  %\IRule% // "unlocked" to "locked" state 
 %\IRule%  %\IRule%  %\IRule% // succ is logically nullptr 
 %\IRule%  %\IRule%  %\IRule% auto R = FetchAndMark() ; 
 %\IRule%  %\IRule%  %\IRule% assert ((uintptr_t(R) & 3) == 0) ; 
 %\IRule%  %\IRule%  %\IRule% if (R == &E) [[likely]] {
 %\IRule%  %\IRule%  %\IRule%  %\IRule% succ = nullptr ; 
 %\IRule%  %\IRule%  %\IRule%  %\IRule% // fast path uncontended arrival ...
 %\IRule%  %\IRule%  %\IRule%  %\IRule% goto EnterCS ; 
 %\IRule%  %\IRule%  %\IRule% } 
 %\IRule%  %\IRule%  %\IRule% // New threads arrived within the exchange-fetch_add() 
 %\IRule%  %\IRule%  %\IRule% // window. 
 %\IRule%  %\IRule%  %\IRule% // As the window is narrow this situation should be 
 %\IRule%  %\IRule%  %\IRule% // relatively rare. 
 %\IRule%  %\IRule%  %\IRule% // We have the onset of contention
 %\IRule%  %\IRule%  %\IRule% // Those new arrivals become our detached entry segment.
 %\IRule%  %\IRule%  %\IRule% // Our &E is now buried at the distal end of the 
 %\IRule%  %\IRule%  %\IRule% // arrival stack
 %\IRule%  %\IRule%  %\IRule% // We need to convey &E thru the chain during succession 
 %\IRule%  %\IRule%  %\IRule% // so that the penultimate thread in the entry segment is
 %\IRule%  %\IRule%  %\IRule% // able to  detect and treat &E as logical end-of-segment 
 %\IRule%  %\IRule%  %\IRule% assert (R%$\rightarrow$%eos.load() == nullptr) ; 
 %\IRule%  %\IRule%  %\IRule% R%$\rightarrow$%eos.store (&E, std::memory_order_release) ; 
 %\IRule%  %\IRule%  %\IRule% succ = R ; 
 %\IRule%  %\IRule% } else { 
 %\IRule%  %\IRule%  %\IRule%  
 %\IRule%  %\IRule%  %\IRule% succ = AnnulMarked (succ) ; 
 %\IRule%  %\IRule%  %\IRule% assert ((uintptr_t(succ) & 3) == 0) ; 
 %\IRule%  %\IRule%  %\IRule% assert (succ %$\neq$% &E) ; 
 %\IRule%  %\IRule%  %\IRule% while (E.Gate.load(std::memory_order_acquire) == 0) { 
 %\IRule%  %\IRule%  %\IRule%  %\IRule% Pause() ; 
 %\IRule%  %\IRule%  %\IRule% }
 %\IRule%  %\IRule%  %\IRule%  
 %\IRule%  %\IRule%  %\IRule% auto eos = E.eos.load (std::memory_order_acquire) ; 
 %\IRule%  %\IRule%  %\IRule% if (eos == nullptr) [[likely]] { 
 %\IRule%  %\IRule%  %\IRule%  %\IRule% goto EnterCS ; 
 %\IRule%  %\IRule%  %\IRule% } 
 %\IRule%  %\IRule%  %\IRule%  
 %\IRule%  %\IRule%  %\IRule% assert (succ %$\neq$% nullptr) ; 
 %\IRule%  %\IRule%  %\IRule%  
 %\IRule%  %\IRule%  %\IRule% // This should be rare
 %\IRule%  %\IRule%  %\IRule% // We only reach here at the onset of contention _AND 
 %\IRule%  %\IRule%  %\IRule% // when the initial arriving thread, which siezed the lock, 
 %\IRule%  %\IRule%  %\IRule% // raced in the exchance()-cas() window above and then 
 %\IRule%  %\IRule%  %\IRule% // became submerged on the arrival stack.  
 %\IRule%  %\IRule%  %\IRule% // Crucially, we usually avoid coherence traffic generated
 %\IRule%  %\IRule%  %\IRule% // by storing into succ%$\rightarrow$%eos
 %\IRule%  %\IRule%  %\IRule%  
 %\IRule%  %\IRule%  %\IRule% // Detect end-of-segment sentinel-marker address
 %\IRule%  %\IRule%  %\IRule% // Annul or squash/quash succ as needed
 %\IRule%  %\IRule%  %\IRule% if (eos == succ) {
 %\IRule%  %\IRule%  %\IRule%  %\IRule% succ = nullptr ; 
 %\IRule%  %\IRule%  %\IRule%  %\IRule% goto EnterCS ; 
 %\IRule%  %\IRule%  %\IRule% }
 %\IRule%  %\IRule%  %\IRule%  
 %\IRule%  %\IRule%  %\IRule% // propapage eos thru chain toward tail 
 %\IRule%  %\IRule%  %\IRule% assert (succ %$\neq$% nullptr) ; 
 %\IRule%  %\IRule%  %\IRule% assert (succ%$\rightarrow$%Gate.load() == 0) ; 
 %\IRule%  %\IRule%  %\IRule% assert (succ%$\rightarrow$%eos.load() == nullptr) ; 
 %\IRule%  %\IRule%  %\IRule% succ%$\rightarrow$%eos.store (eos, std::memory_order_release) ; 
 %\IRule%  %\IRule% } 
 %\IRule%  %\IRule%  
 %\IRule%  %\IRule% // Critical section expressed as lambda 
 %\IRule%  %\IRule% // pass "succ" as context from Acquire to corresponding Release
 %\IRule%  %\IRule% EnterCS: 
 %\IRule%  %\IRule% assert (succ %$\neq$% &E) ; 
 %\IRule%  %\IRule% assert ((uintptr_t(Arrivals.load()) & 2) == 0) ; 
 %\IRule%  %\IRule% assert (uintptr_t(Arrivals.load()) %$\neq$% 0) ; 
 %\IRule%  %\IRule% csfn() ;
 %\IRule%  %\IRule% assert ((uintptr_t(Arrivals.load()) & 2) == 0) ; 
 %\IRule%  %\IRule% assert (uintptr_t(Arrivals.load()) %$\neq$% 0) ; 
 %\IRule%  %\IRule%  
 %\IRule%  %\IRule% // %\underline{Release Phase}% ...
 %\IRule%  %\IRule% // Case : entry list populated 
 %\IRule%  %\IRule% // appoint successor from entry list
 %\IRule%  %\IRule% if (succ %$\neq$% nullptr) {
 %\IRule%  %\IRule%  %\IRule% assert (succ%$\rightarrow$%Gate.load() == 0) ;
 %\IRule%  %\IRule%  %\IRule% succ%$\rightarrow$%Gate.store (1, std::memory_order_release) ; 
 %\IRule%  %\IRule%  %\IRule% return ;
 %\IRule%  %\IRule% }
 %\IRule%  %\IRule%  
 %\IRule%  %\IRule% // Mark MRA element but leave it in emplaced on chain 
 %\IRule%  %\IRule% // FetchAndMark() 
 %\IRule%  %\IRule% // converts locked+arrived to locked+detached
 %\IRule%  %\IRule% // converts locked+detached to unlocked 
 %\IRule%  %\IRule% auto k = FetchAndMark () ; 
 %\IRule%  %\IRule% assert ((uintptr_t(k) & 2) == 0) ; 
 %\IRule%  %\IRule% assert (uintptr_t(k) %$\neq$% 0) ; 
 %\IRule%  %\IRule% if (uintptr_t(k) & 1) return ; 
 %\IRule%  %\IRule%  
 %\IRule%  %\IRule% // We just detached fresh arrivals 
 %\IRule%  %\IRule% assert ((uintptr_t(k) & 3) == 0) ; 
 %\IRule%  %\IRule% assert ((uintptr_t(k) & ~3) %$\neq$% uintptr_t(&E)) ; 
 %\IRule%  %\IRule% assert (k %$\neq$% nullptr) ; 
 %\IRule%  %\IRule% assert (k%$\rightarrow$%Gate.load() == 0) ; 
 %\IRule%  %\IRule% k%$\rightarrow$%Gate.store (1, std::memory_order_release) ; 
 %\IRule% } 
} ;


\end{lstlisting}

\newpage
\lstset{caption={Reciprocating Locks -- Variation}}
\lstset{label={Listing:WFvBD}}
\begin{lstlisting}[mathescape=true,escapechar=\%]
// We observe that this variant is "thread-oblivious" 
// See https://dl.acm.org/doi/10.1145/2370036.2145848
// where one thread can acquire the lock, pass the context to some
// other thread, and that other thread subsquently releases the lock. 
// Special consideration must be given to memory consistency model concerns in this usage,
// and additional fences are typically required.
// A thread-oblivious lock is effectively a binary bounded semaphore
// and that it is trivial to transform such a semaphore into a generalized semaphore. 

struct ReciprocatingLock { 
 %\IRule% struct WaitElement { 
 %\IRule%  %\IRule% std::atomic<int> Gate alignas(128) {0} ; 
 %\IRule%  %\IRule% std::atomic<WaitElement *> eos {nullptr} ; 
 %\IRule% } ; 
 %\IRule%  
 %\IRule% static inline constinit thread_local WaitElement 
 %\IRule%  %\IRule% E alignas(128) {0, nullptr} ; 
 %\IRule% Atomic <WaitElement *> Arrivals {nullptr} ; 
 %\IRule% static inline const auto LOCKEDEMPTY = (WaitElement *) uintptr_t(1) ; 
 %\IRule%  
 public : 
 %\IRule% inline auto %\colorbox{yellow!50}{operator+}%(std::invocable auto && csfn) %$\rightarrow$% void {
 %\IRule%  %\IRule% // %\underline{Acquire Phase}% ...
 %\IRule%  %\IRule% E.eos.store      (nullptr, std::memory_order_release) ; 
 %\IRule%  %\IRule% E.Gate.store     (0      , std::memory_order_release) ; 
 %\IRule%  %\IRule% WaitElement * succ = nullptr ; 
 %\IRule%  %\IRule%  
 %\IRule%  %\IRule% auto tail = Arrivals.exchange (&E) ; 
 %\IRule%  %\IRule% assert (tail %$\neq$% &E) ; 
 %\IRule%  %\IRule% if (tail == nullptr) { 
 %\IRule%  %\IRule%  %\IRule% // fast-path uncontended acquire -- We now hold the lock
 %\IRule%  %\IRule%  %\IRule% // Try to replace &E with SIMPLELOCKED. 
 %\IRule%  %\IRule%  %\IRule% auto R = Arrivals.exchange (LOCKEDEMPTY) ; 
 %\IRule%  %\IRule%  %\IRule% assert (R %$\neq$% nullptr) ; 
 %\IRule%  %\IRule%  %\IRule% if (R %$\neq$% &E) { 
 %\IRule%  %\IRule%  %\IRule%  %\IRule% // We couldn't replace E, but we otherwise detached
 %\IRule%  %\IRule%  %\IRule%  %\IRule% // a new arrival chain, which becomes our entry segment.  
 %\IRule%  %\IRule%  %\IRule%  %\IRule% // This should be rare -- Onset of contention
 %\IRule%  %\IRule%  %\IRule%  %\IRule% // Other threads arrived and pushed onto Arrivals in
 %\IRule%  %\IRule%  %\IRule%  %\IRule% // the exchange-exchange window, above.  
 %\IRule%  %\IRule%  %\IRule%  %\IRule% // Our &E is now buried at distal end of arrival stack
 %\IRule%  %\IRule%  %\IRule%  %\IRule% // The exchange() above snapped off a new entry segment,
 %\IRule%  %\IRule%  %\IRule%  %\IRule% // which is anchored at "R". 
 %\IRule%  %\IRule%  %\IRule%  %\IRule% R%$\rightarrow$%eos.store (&E, std::memory_order_release) ; 
 %\IRule%  %\IRule%  %\IRule%  %\IRule% succ = R ; 
 %\IRule%  %\IRule%  %\IRule% }
 %\IRule%  %\IRule% } else { 
 %\IRule%  %\IRule%  %\IRule% // Coerce LOCKEDEMPTY to nullptr 
 %\IRule%  %\IRule%  %\IRule% // succ will be our successor when we subsequently release
 %\IRule%  %\IRule%  %\IRule% succ = (WaitElement *) (uintptr_t(tail) & ~1) ;
 %\IRule%  %\IRule%  %\IRule% assert (succ %$\neq$% &E) ;
 %\IRule%  %\IRule%  %\IRule%  
 %\IRule%  %\IRule%  %\IRule% // slow path : contention -- waiting phase
 %\IRule%  %\IRule%  %\IRule% while (E.Gate.load (std::memory_order_acquire) == 0) {
 %\IRule%  %\IRule%  %\IRule%  %\IRule% Pause() ; 
 %\IRule%  %\IRule%  %\IRule% }
 %\IRule%  %\IRule%  %\IRule%  
 %\IRule%  %\IRule%  %\IRule% auto eos = E.eos.load(std::memory_order_acquire) ; 
 %\IRule%  %\IRule%  %\IRule% assert (eos %$\neq$% &E) ; 
 %\IRule%  %\IRule%  %\IRule% if (eos %$\neq$% nullptr) [[unlikely]] { 
 %\IRule%  %\IRule%  %\IRule%  %\IRule% // Arriving at this point should be rare. 
 %\IRule%  %\IRule%  %\IRule%  %\IRule% // We only reach here at onset of contention _AND 
 %\IRule%  %\IRule%  %\IRule%  %\IRule% // when the initial arriving thread, which siezed the 
 %\IRule%  %\IRule%  %\IRule%  %\IRule% // lock, raced in the exchance()-exchange() window above 
 %\IRule%  %\IRule%  %\IRule%  %\IRule% // and then became submerged on the arrival stack.  
 %\IRule%  %\IRule%  %\IRule%  %\IRule% // Once in steady-state contention, control will not 
 %\IRule%  %\IRule%  %\IRule%  %\IRule% // reach here. 
 %\IRule%  %\IRule%  %\IRule%  %\IRule% // Crucially, we can then usually avoid coherence 
 %\IRule%  %\IRule%  %\IRule%  %\IRule% // traffic generated by storing into succ%$\rightarrow$%eos. 
 %\IRule%  %\IRule%  %\IRule%  %\IRule%  
 %\IRule%  %\IRule%  %\IRule%  %\IRule% // Detect logical end-of-segment sentinel-marker address
 %\IRule%  %\IRule%  %\IRule%  %\IRule% // Annul or quash succ as needed
 %\IRule%  %\IRule%  %\IRule%  %\IRule% if (eos == succ) {
 %\IRule%  %\IRule%  %\IRule%  %\IRule%  %\IRule% succ = nullptr ; 
 %\IRule%  %\IRule%  %\IRule%  %\IRule%  %\IRule% goto EnterCS ; 
 %\IRule%  %\IRule%  %\IRule%  %\IRule% }
 %\IRule%  %\IRule%  %\IRule%  %\IRule%  
 %\IRule%  %\IRule%  %\IRule%  %\IRule% // propapage eos thru chain toward tail 
 %\IRule%  %\IRule%  %\IRule%  %\IRule% assert (succ %$\neq$% nullptr) ; 
 %\IRule%  %\IRule%  %\IRule%  %\IRule% assert (succ%$\rightarrow$%Gate.load() == 0) ; 
 %\IRule%  %\IRule%  %\IRule%  %\IRule% assert (succ%$\rightarrow$%eos.load() == nullptr) ; 
 %\IRule%  %\IRule%  %\IRule%  %\IRule% succ%$\rightarrow$%eos.store (eos, std::memory_order_release) ; 
 %\IRule%  %\IRule%  %\IRule% }
 %\IRule%  %\IRule% }
 %\IRule%  %\IRule%  
 %\IRule%  %\IRule% // Critical section expressed as lambda 
 %\IRule%  %\IRule% // We pass "succ" as context from Acquire to corresponding Release
 %\IRule%  %\IRule% EnterCS: 
 %\IRule%  %\IRule% assert (Arrivals.load() %$\neq$% nullptr) ; 
 %\IRule%  %\IRule% csfn() ;
 %\IRule%  %\IRule% assert (Arrivals.load() %$\neq$% nullptr) ; 
 %\IRule%  %\IRule%  
 %\IRule%  %\IRule% // %\underline{Release Phase}% ...
 %\IRule%  %\IRule% if (succ == nullptr) { 
 %\IRule%  %\IRule%  %\IRule% // Case : entry list and arrivals are both empty
 %\IRule%  %\IRule%  %\IRule% // try fast-path uncontended unlock 
 %\IRule%  %\IRule%  %\IRule% auto v = Arrivals.cas (LOCKEDEMPTY, nullptr) ;
 %\IRule%  %\IRule%  %\IRule% assert (v %$\neq$% nullptr && v %$\neq$% &E) ; 
 %\IRule%  %\IRule%  %\IRule% if (v == LOCKEDEMPTY) return ;
 %\IRule%  %\IRule%  %\IRule%  
 %\IRule%  %\IRule%  %\IRule% // Detach a new arrival segment
 %\IRule%  %\IRule%  %\IRule% succ = Arrivals.exchange (LOCKEDEMPTY) ;
 %\IRule%  %\IRule%  %\IRule% assert (succ %$\neq$% nullptr && succ %$\neq$% LOCKEDEMPTY) ; 
 %\IRule%  %\IRule% } 
 %\IRule%  %\IRule% assert (succ%$\rightarrow$%Gate.load() == 0) ; 
 %\IRule%  %\IRule% succ%$\rightarrow$%Gate.store (1, std::memory_order_release) ;
 %\IRule% }
} ; 




\end{lstlisting}

\twocolumn

\newpage 

\section{Retrograde Ticket Lock} 

In Listing-\ref{Listing:TKT-Retrograde} we describe the \emph{retrograde ticket lock} 
algorithm, which mimics the admission order policy of \recipro{} but is implemented
as a version of the classic ticket lock algorithm.

\FloatBarrier 

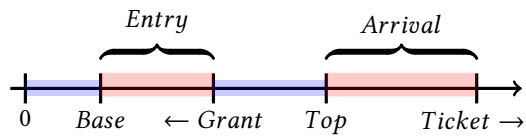
\begin{figure}[h]  
\begin{tikzpicture}[mydrawstyle/.style={draw=black, very thick}, x=1mm, y=1mm, z=1mm]
  \draw[mydrawstyle, ->](-2,30)--(66,30) node at (-6,30)[left]{};
  \draw[mydrawstyle](0,28)--(0,32) node[below=10]{$0$};
  \draw[mydrawstyle](10,28)--(10,32) node[below=10]{$Base$};
  \draw[mydrawstyle](25,28)--(25,32) node[below=10]{$\leftarrow{}Grant$};
  \draw[mydrawstyle](40,28)--(40,32) node[below=10]{$Top$};
  \draw[mydrawstyle](60,28)--(60,32) node[below=10]{$Ticket\rightarrow{}$};
  \fill[fill=blue, opacity=0.2](0,29)--(0,31)--(10,31)--(10,29)--(0,29);
  \fill[fill=blue, opacity=0.2](25,29)--(25,31)--(40,31)--(40,29)--(25,29);
  \filldraw[draw=red,fill=red,opacity=0.2] (10,29) rectangle (25,32);
  \filldraw[draw=red,fill=red,opacity=0.2] (40,29) rectangle (60,32);
  \draw[decorate,decoration={calligraphic brace,amplitude=5pt},line width=2.25pt] (10,34) -- (25,34) node [midway,above=5] {$Entry$};
  \draw[decorate,decoration={calligraphic brace,amplitude=5pt},line width=2.25pt] (40,34) -- (60,34) node [midway,above=5] {$Arrival$};
\end{tikzpicture}
\caption{Retrograde Ticket Locks in Action} 
\label{retrograde} 
\end{figure} 

The classic ticket lock uses \texttt{grant} and \texttt{ticket} fields, where arriving
threads atomically fetch-and-increment \texttt{ticket} and then wait for the assigned ticket value to equal \texttt{grant}
and the corresponding \Release{} operator increments \texttt{grant}.  
For retrograde ticket locks we add new per-lock \texttt{base} and \texttt{top} fields. 
In Figure-\ref{retrograde} we show a number line depicting the trajectory of the lock fields. 
Assigned tickets in the range $\texttt{base}-\texttt{top}$ represent the entry segment and
those in range $\texttt{top}-\texttt{ticket}$ represent the arrival segment. 
Regions colored blue represent ticket values already been granted ownership and no longer waiting, 
while those colored red are waiting for admission.  
Crucially, while the entry segment remains populated with waiting threads, the \Release{}
operator advances the \texttt{grant} field in a descending fashion, yielding a \emph{retrograde} order.
When \texttt{grant} reaches \texttt{base}, the entry segment is exhausted, and
we reprovision the entry segment by setting \texttt{base} to \texttt{top} and \texttt{top} to \texttt{ticket}.  
That is, the arrival segment becomes the new entry segment.  
By convention, \texttt{top} and \texttt{base} 
are accessed only by the current lock holder and only within the \Release{} operation.

As we use magnitude-based comparisons, arithmetic roll-over and aliasing of 
the \texttt{ticket} and \texttt{grant} becomes a concern.
To address that issue we simply ensure that \texttt{ticket}-related fields are 64-bit integers.
Assuming a processor could increment a value at most once per nanosecond, these fields would not overflow
and wrap around in less than 200 years, so arithmetic overflow is not a practical concern.
Assuming that \texttt{std::atomic<int64\_t>:fetch\_add(1)} is constant-time -- which 
depends on both the library implementation of \texttt{std::atomic} and on the platform capabilities --
our \emph{doorway} phase is also also constant-time.   Only one atomic read-modify-write \texttt{fetch\_add} 
operation is required in \Acquire{} and none in \Release{}, and \Release{} runs in constant-time.  

We are not constrained to simple retrograde admission order within the entry segment.  
Lets say our entry segment consists of threads $A-B-C-D$ which hold ticket values $1005-1004-1003-1002$, respectively.
$A$ is the most recently arrived thread in the entry segment.  
Each thread waits for its corresponding ticket value
to appear in the \texttt{Grant} field, which confers and conveys ownership.  For retrograde admission, which mimics
\recipro{}, our admission order is $A$ then $B$ then $C$ then $D$, using descending ticket values with the entry segment.   
A \emph{prograde} admission order of $D$ then $C$ then $B$ then $A$, with ascending ticket values, is tantamount to
simple classic FIFO ticket locks.   
By using ticket values, we are able, in constant-time, to activate and enable a thread at any \emph{arbitrary} position (offset) 
in the entry segment, unlike \recipro{} where the order is dictated, as threads know only the identity of their immediate neighbor.  
Using ticket-based succession allows a wider variety of admission orders, compared to \recipro{}, as
ticket-base successor allows random access to the elements of the entry segment and thus more latitude for
succession order.

\Invisible{MRAT = most recently arrived thread}

For example, a viable approach is the following.  In the \Release{} operator we run a biased 
Bernoulli trial, and, based on the outcome, select the successor from either the head of the 
remaining entry segment -- the most-recently arrived thread -- or the tail, which is the least
recently arrived thread, but with the probability favoring the head.  
This yields a mostly LIFO admission order with the entry segment -- mostly retrograde, but occasionally prograde.  
Crucially, we use the tunable Bernoulli probability to strike a balance between 
fairness over a period, and aggregate throughput.  
Such randomization is sufficient to break or perturb long-term unfairness arising
from repeating palindromic admission cycles.

As an optimization, to reduce the use of the random number generator, we can implement a
per-lock \texttt{CountDown} variable.  The \Release{} operator always decrements the counter.  
If the value is found greater than zero, we extract a successor from the head of the entry segment.
Otherwise, we extract a successor from the tail, compute a small uniform random integer in the range $1..M$ and
then reset \texttt{CountDown} to that value.  
For our purposes, a simple low-latency low-quality pseudo-random generator number suffices, such as a single-word
Marsaglia \emph{xor-shift} variant\cite{jss03-marsaglia}.  

A related technique is, in \Release{}, when detaching a new arrival segment, to
run a Bernoulli trial to pick a succession direction -- prograde or retrograde but biased toward and
favoring retrograde -- and then
use that direction for the entirety of the segment.  

These forms use randomization to provide long-term statistical avoidance of both unfair admission and unfair 
cache residency, but still provide better aggregate cache residency (and throughput) than simple FIFO
while also retaining the desirable thread-specific bounded bypass property.

For all the above, we can also borrow the waiting scheme from TWA and construct versions 
that avoid global spinning. 

\Invisible{strikes compromise; trade-off; balance; tension}  
\Invisible{derangement, disorder} 
\Invisible{Pick random starting position and direction}  

\Invisible{We conjecture that retrograde palindromic order may also provide innate NUMA benefits, reducing
\emph{lock migration} by virtue of the palindromic schedule.  If we conceptualize NUMA affinity as a degenerate
cache of just one element, then as a corollary to improved cache residency, we also reduce lock migration
rates via a palindromic schedule.} 

\FloatBarrier

\onecolumn
\lstset{caption={Retrograde Ticket Locks}}
\lstset{label={Listing:TKT-Retrograde}}
\begin{lstlisting}[mathescape=true,escapechar=\%]
struct TicketLock { 
 %\IRule% std::atomic <int> Ticket {0} ; 
 %\IRule% std::atomic <int> Grant  {0} ; 
} ; 

// Classic Ticket lock ...

void %\colorbox{yellow!50}{TicketAcquire}% (TicketLock * L) {
 %\IRule% auto ticket = L%$\rightarrow$%Ticket.fetch_add(1) ; 
 %\IRule% while (L%$\rightarrow$%Grant %$\neq$% ticket) {
 %\IRule%  %\IRule% Pause() ; 
 %\IRule% }
}

void %\colorbox{yellow!50}{TicketRelease}% (TicketLock * L) { 
 %\IRule% auto grant = L%$\rightarrow$%Grant.load() + 1 ; 
 %\IRule% L%$\rightarrow$%Grant.store (grant, std::memory_order_release) ; 
} 

// "Retrograde" ticket lock 
// provides the same admission order as reciprocating locks
// Note that the Acquire() operator is identical to that of 
// the baseline ticket lock, above.  
// All changes relative to ticket locks are encapsulated
// in the unlock operator. 
// 
// Invariant : Ticket >= Top >= Grant >= Base
// Threads with tickets values in [Top,Base] are admitted in reverse order
// Waiting   : Ticket thru Top union with Grant thru Base
// Completed : Top thru Grant union with those < Base 
// EntrySet  : Grant thru Base
// Arrivals  : Ticket thru Top
// Note that only the current lock hold accesses Top and Base.  
// In particular, the lock itself protects Top and Base.


struct RetrogradeLock { 
 %\IRule% std::atomic <int64_t> Ticket {0} ; 
 %\IRule% std::atomic <int64_t> Grant  {0} ; 
 %\IRule% int64_t Top  = 0 ;        
 %\IRule% int64_t Base = 0 ;  
} ; 

void %\colorbox{yellow!50}{RetrogradeAcquire}% (RetrogradeLock * L) { 
 %\IRule% auto ticket = L%$\rightarrow$%Ticket.fetch_add(1) ; 
 %\IRule% while (L%$\rightarrow$%Grant %$\neq$% ticket) {
 %\IRule%  %\IRule% Pause() ; 
 %\IRule% }
}

void %\colorbox{yellow!50}{RetrogradeRelease}% (RetrogradeLock * L) { 
 %\IRule% auto grant = L%$\rightarrow$%Grant.load() - 1 
 %\IRule% if (grant > L%$\rightarrow$%Base) { 
 %\IRule%  %\IRule% // Region of reverse admission order ...
 %\IRule%  %\IRule% // working our way backward through entry set 
 %\IRule%  %\IRule% L%$\rightarrow$%Grant = grant ; 
 %\IRule%  %\IRule% return ; 
 %\IRule% } 
 %\IRule% auto Hi  = L%$\rightarrow$%Top ;
 %\IRule% L%$\rightarrow$%Base  = Hi ; 
 %\IRule% auto tmp = L%$\rightarrow$%Ticket.load() ; 
 %\IRule% L%$\rightarrow$%Top   = tmp - 1 ; 
 %\IRule% if (tmp == (Hi + 1)) { 
 %\IRule%  %\IRule% // Apparently no waiters
 %\IRule%  %\IRule% // Uncontended release -- set to "unlocked" state with 
 %\IRule%  %\IRule% // Ticket == Grant.
 %\IRule%  %\IRule% // Benign if Ticket advances concurrently after we read it.
 %\IRule%  %\IRule% // The store into L%$\rightarrow$%Grant must become visible after the 
 %\IRule%  %\IRule% // other stores. 
 %\IRule%  %\IRule% L%$\rightarrow$%Top   = tmp ; 
 %\IRule%  %\IRule% L%$\rightarrow$%Base  = tmp ; 
 %\IRule%  %\IRule% L%$\rightarrow$%Grant = tmp ;
 %\IRule% } else {
 %\IRule%  %\IRule% // Contention with waiters -- stays locked
 %\IRule%  %\IRule% L%$\rightarrow$%Grant = tmp - 1 ;
 %\IRule% }
} 


\end{lstlisting}
\twocolumn 

\newpage
\onecolumn 
\section{Alternative ``Gated'' Formulation} 
\FloatBarrier
\lstset{caption={Reciprocating Locks -- Alternative ``Gated'' Formulation}}
\lstset{label={Listing:CPSGatedBN}}
\lstinputlisting[mathescape=true,escapechar=\%]{listing-CPSGatedBN.cc-fx}
\twocolumn 

\newpage
\onecolumn 
\section{Alternative ``2 Lanes'' Formulation -- Imposes Long-Term Fairness} 
\FloatBarrier
\lstset{caption={Reciprocating Locks -- Alternative ``2 Lanes'' Formulation -- Imposes Long-Term Fairness}}
\lstset{label={Listing:CPSGated2XLanes}}
\lstinputlisting[mathescape=true,escapechar=\%]{listing-CPSGated2XLanes.cc-fx}
\twocolumn 
\end{document}